\documentclass[prd,aps,twocolumn,a4paper,showkeys,nofootinbib]{revtex4-1}

\usepackage{amsmath}
\usepackage{amsfonts}
\usepackage{amssymb}	
\usepackage{graphicx}
\usepackage{amsbsy}
\usepackage{color}
\usepackage{commath}
\usepackage{hyperref}
\usepackage[normalem]{ulem}

\allowdisplaybreaks

\def\d{\mathrm{d}}
\newcommand{\rmi}{\mathrm{i}}
\newcommand{\rme}{\operatorname{e}}

\begin{document}

\title{A numerical-relativity surrogate model for hyperbolic encounters of black holes: challenges in parameter estimation}

%\title{An NRSurrogate for hyperbolic encounters of black holes}

\author{Joan \surname{Fontbut\'e}${}^{1,2}$}
\author{Tomas \surname{Andrade}${}^{1}$}
\author{Raimon \surname{Luna}${}^{3,4}$}
\author{Juan \surname{Calder\'on Bustillo}${}^{5,6}$}
\author{Gonzalo \surname{Morr\'as}${}^{7}$}
\author{Santiago \surname{Jaraba}${}^{7}$}
\author{Juan \surname{Garc\'ia-Bellido}${}^{7}$}
\author{Germán \surname{L\'opez Izquierdo}${}^{1}$}

\affiliation{${}^1$ Departament de F{\'\i}sica Qu\`antica i Astrof\'{\i}sica, Institut de Ci\`encies del Cosmos, Universitat de Barcelona, Mart\'{\i} i Franqu\`es 1, E-08028 Barcelona, Spain}

\affiliation{${}^2$ Theoretisch-Physikalisches Institut, Friedrich-Schiller-Universit{\"a}t Jena, 07743, Jena, Germany}  

\affiliation{${}^3$ Departamento de Astronom\'{i}a y Astrof\'{i}sica, Universitat de Val\`{e}ncia, Dr. Moliner 50, 46100, Burjassot (Val\`{e}ncia), Spain}

\affiliation{${}^4$ Departamento de Matem\'{a}tica da Universidade de Aveiro and Centre for Research and Development in Mathematics and Applications (CIDMA), Campus de Santiago, 3810-193 Aveiro, Portugal}

\affiliation{${}^5$ Instituto Galego de F\'{i}sica de Altas Enerx\'{i}as, Universidade de Santiago de Compostela, 15782 Santiago de Compostela, Galicia, Spain}

\affiliation{${}^6$ Department of Physics, The Chinese University of Hong Kong, Shatin, N.T., Hong Kong}

\affiliation{${}^7$ Instituto de F\'isica Te\'orica IFT-UAM/CSIC, Universidad Aut\'onoma de Madrid, Cantoblanco 28049 Madrid, Spain} 

\begin{abstract}
    We present a surrogate numerical-relativity model for close hyperbolic black-hole encounters with equal masses and spins aligned with the orbital momentum. Our model, generated in terms of the Newman-Penrose scalar $\psi_4$, spans impact parameters $b/M\in [11, 15]$ and spin components $\chi_{i} \in [-0.5,0.5]$, modeling the $(\ell,m)=(2,0)$, $(2, \pm 2)$, $(3,\pm 2)$ and $(4,\pm 4)$ emission multipoles. The model is faithful to numerical relativity simulations, yielding mismatches lower than $10^{-3}$. We test the ability of our model to recover the parameters of numerically simulated signals. We find that, despite the high accuracy of the model, parameter inference struggles to correctly capture the parameters of the source even for SNRs as large as 50 due to the strong degeneracies present in the parameter space. This indicates that correctly identifying these systems will require of extremely large signal loudness, only typical of third generation detectors. Nevertheless, we also find that, if one attempts to infer certain combinations of such degenerated parameters, there might be a chance to prove the existence of this type of events, even with the current ground-based detectors, as long as these combinations make sense astrophysically and cosmologically.

\end{abstract}

\date{\today}

\maketitle

\section{Introduction}

%\jgb{I am working here...}

Since the detection of the black hole binary GW150914 \cite{LIGOScientific:2016aoc}, gravitational wave (GW) astronomy has opened up the possibility of directly studying black holes (BHs). To date, the LIGO-Virgo-KAGRA (LVK) collaboration has detected 90 compact binary coalescence (CBC) events, predominantly black hole binaries \cite{Abbott2019_GWTC1,Abbott2021_GWTC2,Abbott2023_GWTC3}. 
Most of these have been classified as quasi-circular, indicative of black holes binaries evolving as isolated systems that subsequently circularize due to the emission of GWs.

As detector sensitivity continues to improve, the scope for detecting novel events expands. Among these, considerable attention has been devoted to scenarios involving dense environments \cite{CanevaSantoro2024,Leong2023, Mukherjee:2020hnm,Cho:2018upo,Dandapat:2023zzn}, such as globular clusters, active galactic nucleus (AGNs) \cite{Ford2022,Graham2020,Graham2023}, and clusters of primordial black holes \cite{Clesse:2016vqa,Trashorras:2020mwn,Siles:2024yym,Carr:2023tpt}. 
The high density of these environments facilitates the occurrence of close encounters between black holes, resulting in the emission of bursts of GW radiation, a phenomenon referred to as close hyperbolic encounters (CHE)~\cite{Garcia-Bellido:2017knh,Garcia-Bellido:2017qal,Morras:2021atg,Jaraba:2021ces,Garcia-Bellido:2021jlq,Caldarola:2023ipo}.

The observation of stellar-mass CHE events is within the frequency band accessible to current ground-based detectors \cite{Garcia-Bellido:2017qal,Morras:2021atg,Dandapat:2023zzn}. However, while the search for such events within LVK data has been undertaken \cite{Morras:2021atg,Bini:2023gaj}, no CHE has been detected so far. Nevertheless, there is the intriguing prospect of detecting CHE encounters involving black holes of primordial origin \cite{Garcia-Bellido:2017knh}, potentially within the capabilities of A+ and future interferometers like Einstein Telescope \cite{Punturo:2010zz} and Cosmic Explorer \cite{Reitze:2019iox}. 

If the GWs emitted during a CHE are sufficiently strong, the system can become bound, undergoing a so-called dynamical capture~\cite{Gold:2012tk,Nagar:2020xsk,Andrade:2023trh,Albanesi:2024xus}. 
These can entail a series of CHE before merger, displaying a complex waveform phenomenology. The most massive event detected to date, GW190521, has been proposed to be a dynamical capture with one previous close passage~\cite{Gamba:2021gap}, although such passage was not searched for in the data. 

Models used for searches of CHE in \cite{Morras:2021atg}, \cite{Bini:2023gaj} were based on 3.5 post-Newtonian (PN) approximations. On general grounds, the use of PN is justified when the velocities are small and BH separations are large, but detailed quantitative comparisons with Numerical Relativity (NR) simulations of the PN waveforms are lacking, as well as a precise study on the validity of the PN approximation for CHE. 
As a circumstantial evidence of PN performance, we can mention that the PN/NR comparisons of the scattering angles carried out with lower-order PN models have shown to be poor~\cite{Damour:2014afa}. 

Another approximate scheme that has been employed to describe CHE is the Post-Minkowskian (PM) approximation. This is complementary to PN, and extends the regime of validity allowing for higher velocities although remaining in the weak-field limit. 
Recent studies show that naive PM approximations display poor convergence to NR data in the spinless and spinning cases. It is only after carrying out resummation techniques employed in the effective one-body (EOB) formalism that the comparison between PM and NR scattering angles is satisfactory~\cite{Rettegno:2019tzh,Damour:2022ybd}.

NR simulations for CHE have been carried out in~\cite{Damour:2014afa,Nelson:2019czq,Jaraba:2021ces,Rettegno:2023ghr,Albanesi:2024xus}, but none of them are publicly available. Most of the Physics that has been extracted from these simulations are the initial and final properties, together with scattering angles. A technical difficulty is that open source NR codes such as \texttt{EinsteinToolkit} \cite{EinsteinToolkit:2024_05,Loffler:2011ay,EinsteinToolkit} allow for the direct extraction of $\psi_4$ at finite radius, which requires post-processing to obtain the strain \cite{Hopper:2022rwo}. 
In particular, the fixed-frequency integration technique employed to recover the strain for quasi-circular (QC) binaries \cite{Reisswig:2010di} does not immediately extend to CHE or dynamical captures due to the lack of a well motivated cut-off frequency \cite{Andrade:2023trh}, so one cannot robustly bypass the unphysical drift that arises when performing two consecutive time integrations of the Weyl scalar $\psi_4$. 

Despite the aforementioned issues, the current PN models of CHE are likely to suffice for detection purposes, specially when integrated in weakly modeled searches such as the Coherent WaveBurst (cWB) pipeline \cite{Drago:2020kic}. On the contrary, proper parameter estimation of upcoming events of this kind will require model improvements and NR validation.
Motivated by the difficulties with the approximate PN models to capture NR data, and trying to ameliorate the numerical artefacts introduced by numerical time integration, we tackle the construction of an NR surrogate model for CHE built upon the $\psi_4$-scalar modes, which can also be used to analyse GW data avoiding the obtention of the GW strain \cite{CalderonBustillo:2022dph,CaldernBustillo2023_proca2,Luna:2024kof}.
The relative simplicity of the waveforms suggests that this approach should be efficient, requiring a relatively sparse coverage of parameter space compared to, say, QC binaries. Nevertheless, the parameter space for QC binaries has one less dimension. Therefore, covering the parameter space for more generic orbits can be far from trivial.  

As a proof of principle, in this paper we consider a surrogate model built out of a set of NR simulations of CHE at fixed energy $E = 1.023 M$ and mass ratio $q = 1$, with impact parameter $\hat{b}$ varying in $\hat{b}=b/M \in [11, 15]$ and equal spins in $\chi \in [-0.5, 0.5]$. 
Covering this 2D parameter space with a grid of 63 roughly equi-spaced NR simulations detailed in Sec. \ref{sec:NR}, we show in Sec. \ref{sec:surr} how our NR surrogate achieves an average unfaithfulness of $10^{-3}$ over a dynamically chosen test set in k-fold cross-validation. 

After validating the model, in Sec. \ref{sec:injections} we then use it to recover NR simulations injected in zero-noise, using the power spectral density of Advanced LIGO \cite{TheLIGOScientific:2014jea} and Advanced Virgo \cite{VIRGO:2014yos} detectors at the time of the GW190521 event \cite{GW190521D,GWOSC_GW190521}. We consider a true detector-frame total mass of $M=60 M_\odot$ and two different loudness characterised by the optimal signal-to-noise ratio (SNR) of the injection, namely $\sim 15$ (typical of current events) and $\sim 50$ (roughly twice the loudest SNR to date). These SNRs correspond to luminosity distances $d_L\in [130,440]$ Mpc and $d_L\in[40,135]$ Mpc respectively. We find strong degeneracies in the parameter space, understood as different parameters (e.g. impact parameter, total mass or luminosity distance) having a very similar impact in the waveform, which prevent accurate estimates of the source parameters even at large SNRs. 

%where the intrinsic parameters (e.g impact parameter, spin, source-frame total mass) mix with the extrinsic parameters (e.g luminosity distance and detector-frame total mass). 
% This is largely due to the simplicity of the waveforms, which becomes apparent through the dimensionality reduction we employ before interpolating the data. In particular, as we shall see, Singular Value Decomposition of the time domain waveforms 
 
The rest of this paper is organised as follows. In Sec. \ref{sec:OP}, we perform the analytical derivation of the leading order PN waveforms in the frequency domain for hyperbolic orbits, finding a pair of parameters which are less affected by the aforementioned degeneracies. These combinations, given in
\eqref{eq:PN_parameters_surr:l1}-\eqref{eq:PN_parameters_surr:l2} below,  approximately correspond to the overall amplitude and the peak frequency of the strain. 
We verify that, in fact, the posteriors of these parameters are much more sharply defined when performing NR parameter recovery, as well as the proof that if one manually breaks the degeneracy when performing the analysis, the intrinsic parameters (e.g impact parameter) are recovered satisfactorily.

\section{NR simulations \label{sec:NR}}

\subsection{Setting}

We have carried out a set of 63 NR simulations of CHE sources using the open source code \texttt{EinsteinToolkit} \cite{EinsteinToolkit:2024_05,Loffler:2011ay,EinsteinToolkit}. Our setup follows that one used in \cite{Andrade:2023trh}. 
In particular, we use the thorn \texttt{TwoPunctures}~\cite{Brandt:1997tf, Ansorg:2004ds} to solve for the initial data and \texttt{MLBSSN}, which implements the BSSN formulation of General Relativity \cite{Baumgarte:1998te, Shibata:1995we} for time evolution. We take the usual choices for the initial lapse $\alpha_{0}=\Psi_{\rm BH}^{-2}$ and the initial shift $\beta^i_0=0$.
We use 8th order finite differences for spatial derivatives and a method-of-lines time integration with a 4th order Runge-Kutta scheme. 
We employ Kreiss-Oliger dissipation of order 9 scaled with a factor of $\epsilon= 0.1$, and a CFL factor of $0.1$.
Our computational grid is a box with all edges having length $L = 640 M$, with the coarsest level having resolution $dx = 4M$. We use 
Berger-Oliger mesh refinement with 9 box-in-box levels, with the two most refined boxes containing the two punctures, so that resolution at puncture is equal to $dx_p = 4/2^8M = 0.015625M.$
We employ z-reflection symmetry to reduce the computational domain.  

We extract the wave content of our simulations using the thorns \texttt{WeylScal4} and \texttt{Multipole} which output the Weyl scalar $\psi_4$ expanded in spherical harmonics up to $l = 4$, at fixed radius $R$. 

We have checked that $R \psi_4$ changes by less than $0.1\%$  as we vary $R$ in $\{70, 80, 90, 100, 110\} M$, which are all located in a refinement level with resolution $dx = 4/2^2M = 1M$. Unless otherwise stated, we use the waveforms extracted at finite radius $R_{\rm ext} = 100 M$.

\subsection{Initial Data}

We follow and expand on the setup \cite{Damour:2014afa}. More specifically, the initial black holes are located on the x-axis with positions $\pm X$, and initial momenta 
%\gmc{Is there a $1/(2X)$ missing in $P_y$?}
\begin{equation}
    (P_x, P_y, P_z) = \pm P( - [1 - (b/2X)^2 ]^{1/2}, b/(2X), 0),
    \label{eq:P_initial}
\end{equation}
\noindent where $b$ is the impact parameter, related to the initial ADM angular momentum by $J_{\text{ADM}} = 2 X |P_y| = P b$. We also introduce rotation in the initial black holes aligned with the orbital angular momentum as $\vec J_{1, 2} = \chi M_{1, 2}^2 (0, 0, 1)$. Here $M_{1,2}$ are the puncture ADM masses and $\chi$ the rotation parameter. The total mass of the simulation is given by $M = M_1 + M_2$. 
Our initial data is $z-$reflection symmetric, which allows us to reduce the cost of the simulations.

Throughout this work we restrict ourselves to $P = 0.11456439 M$, $q = M_2/M_1 = 1$, and $X = 50 M$, so that the initial separation is $D = 100 M$. This coincides with the setup in \cite{Damour:2014afa}, which only considered non-spinning initial data, which eases the comparison with some of their simulations. 
Our initial data is then described by two parameters, $\hat{b}$ and $\chi$. We consider values such that $\hat{b} \in [11, 15]$ and $\chi \in [-0.5, 0.5]$, as shown in Fig. \ref{fig:param_space}. 
For values of the impact parameter $\hat{b} < 11.0$ and negative spins, we find configurations that end in merger. These will be reported elsewhere. 

\subsection{Simulation results}

\begin{figure}[thpb] 
\centering
\includegraphics[width=0.45\textwidth]{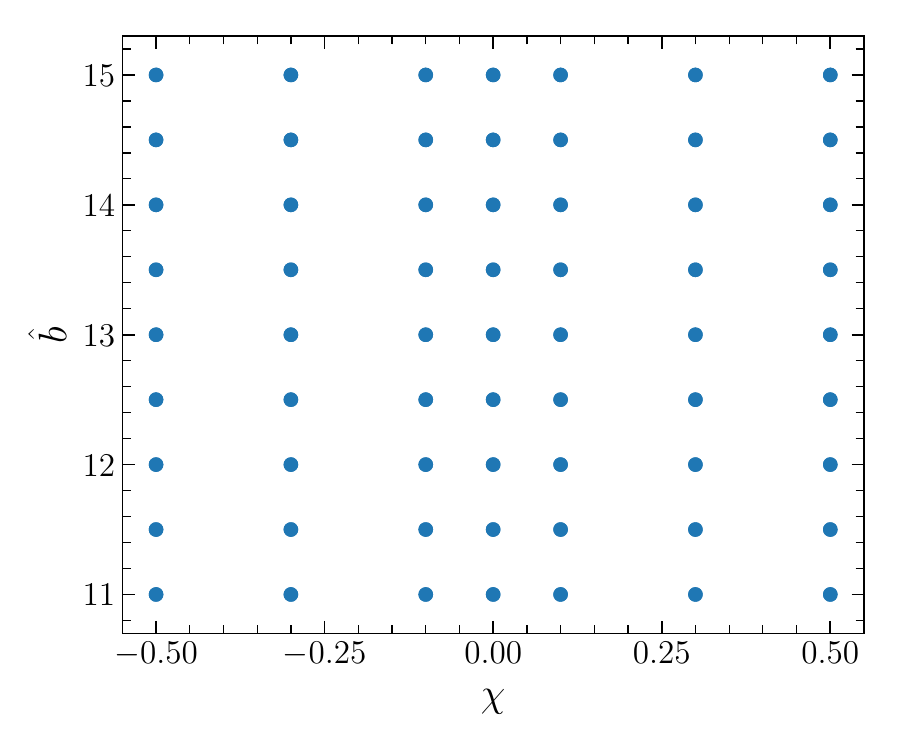}
\caption{Scheme showing the set of 63 simulations used to cover our parameter space.} 
\label{fig:param_space}
\end{figure}

\begin{figure}[htpb]
\begin{center}
\includegraphics[width=0.45\textwidth]{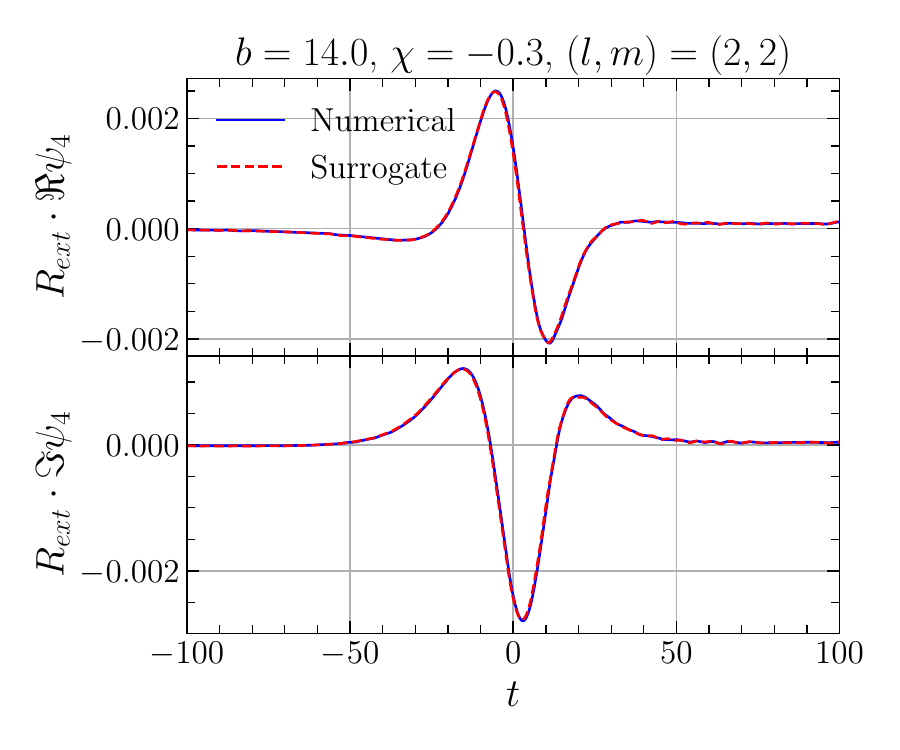}
\caption{Example of a numerical waveform from the test set, together with its prediction by the surrogate model.
\label{fig:wave_example}}
\end{center}
\end{figure}

We show in Fig. \ref{fig:param_space} the parameter space covered by our simulations. For the range of parameters we have studied, all energies are approximately $E_{\text{ADM}} \approx 1.023 M$, and all trajectories are open orbits. 
We show the leading mode of $\psi_4$ for a specific value of the initial parameters in Fig. \ref{fig:wave_example}.
We note that the leading mode $(2,2)$ dominates over the next sub-leading modes $(2,0)$, $(3,2)$, $(4,4)$ by a factor of $\sim$ 8, 20 and 5 respectively, which is expected for equal mass ratios in such faced-on configurations with the motion restricted to the x-y plane. %\jc{Actually, I remember the 20 mode being almost co-dominant for head-on mergers}. 

In addition, the results for the scattering angles from the complete set of simulations shown in Fig. \ref{fig:param_space} are presented in Appendix \ref{app:spin-effects}.

\subsection{Consistency checks}

Our configurations with $\chi = 0$ have previously been considered in \cite{Damour:2014afa}. We find good agreement both qualitatively e.g. in the shape of the trajectories, and quantitatively, e.g the scattering angles coincide within less than $0.7 \%$. 
In addition, we have carried out simulations with $\hat{b} = 11$, $\chi = 0$ at varying resolutions finding an approximate convergence of order 6 for the leading $(2,2)$ mode of the $\psi_4$ scalar. 
See Appendix \ref{app:consistency} for more details on our consistency checks. 

\section{NR Surrogate \label{sec:surr}}

\begin{figure}[htpb]
\begin{center}
\includegraphics[width=0.49\textwidth]{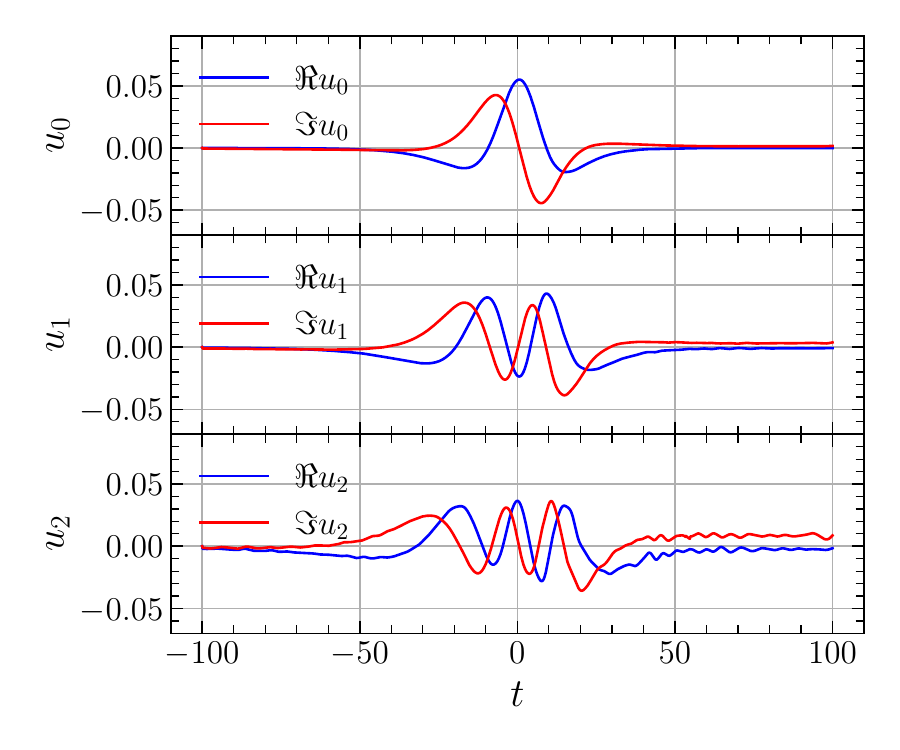}
\caption{The first vectors in the SVD reduced basis of the surrogate.
\label{fig:basis_vectors}}
\end{center}
\end{figure}

The numerical relativity surrogate model is based on a two-stage scheme as detailed in \cite{Luna:2024kof}. First, a dimensionality reduction by singular value decomposition (SVD) is applied on the training dataset of waveforms. This allows us to encode a general waveform in a small number of complex values, namely the coefficients of its expression in terms of a reduced vector basis. Indeed, by taking the $n$ training complex waveforms of length $m$ as columns of an $m \times n$ matrix $M$, we can decompose it as 
\begin{equation}
    M = U \Sigma V^\dagger\; .
\end{equation}
where $U$ and $V$ are $m \times m$ and $n \times n$ unitary matrices, respectively. $\Sigma$ is an $m \times n$ diagonal matrix with non-negative real diagonal elements, which are known as the singular values of the system. The columns of $U$ are orthonormal and can be taken as a basis for the description of the dataset. Each successive basis vector has a lower importance on the description of the data, quantified by its associated singular value, than the preceding ones. Therefore, the basis can be truncated once the desired accuracy is achieved. The first three vectors of the surrogate reduced basis are shown in Fig. \ref{fig:basis_vectors}.
The complex coefficients of each waveform in terms of the SVD reduced basis depend on the physical parameters $(\hat{b}, \chi)$ and have to be interpolated accordingly to complete the surrogate. This is performed under the \texttt{CloughTocher2DInterpolator} algorithm from \texttt{scipy}, which triangulates the input data over a convex hull with \texttt{Qhull} \cite{QHull} and then uses a Clough-Tocher scheme to define cubic B\'ezier surfaces on the triangles \cite{Clough-Tocher}. 

\begin{figure}[htpb]
\begin{center}
\includegraphics[width=0.45\textwidth]{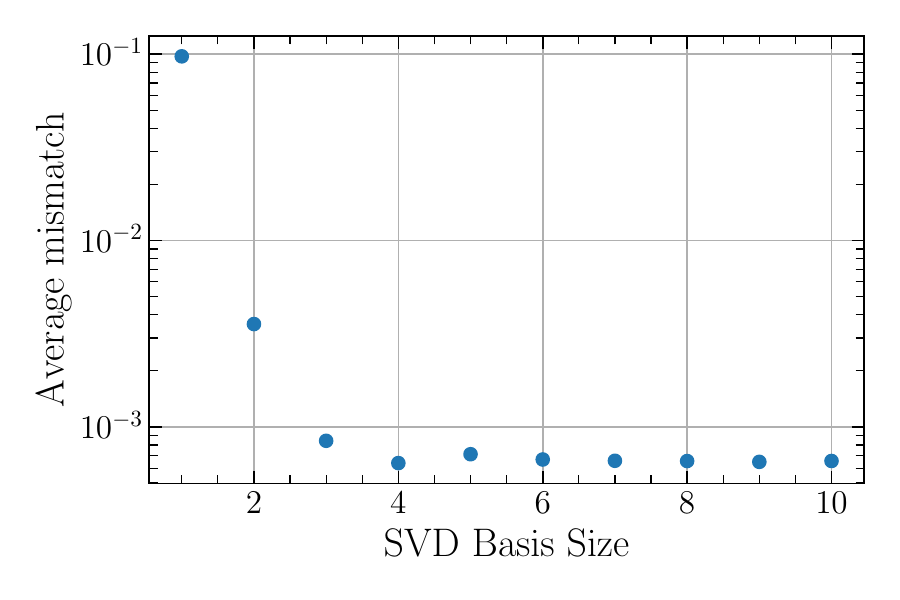}
\caption{Average mismatch for different dimensions of the reduced basis size, evaluated by $K$-folding on the training set. The quality of the surrogate seems to remain constant if the dimension of the basis is 3 or larger. 
\label{fig:basis_size}}
\end{center}
\end{figure}

The selection of the model hyperparameters, in this case the dimension of the SVD basis, is typically performed on a validation set in order to avoid overfitting on the testing set. However, as our full dataset has a very limited number of samples, we implemented a $K$-folding cross validation scheme (see \cite{Luna:2024kof} for details) in which the training and validation roles are rotated over $K = 5$ folds of the training set. This scheme allows us to use all the training data to select the best model type. Fig. \ref{fig:basis_size} shows the average mismatch of the training set waveforms for surrogate models with different numbers of basis vectors. The mismatch between signals is defined as \cite{Cutler:1994ys, Apostolatos:1995pj,Finn:1992wt}:
\begin{equation}
    \mathcal{\overline{F}}\equiv 1-\mathcal{F}=1-\max_{t_s,\phi}\frac{(\psi_{4_1}|\psi_{4_2})}{\sqrt{(\psi_{4_1}|\psi_{4_1})(\psi_{4_2}|\psi_{4_2})}} \label{eq:mism},
\end{equation}
where the inner-products $(\psi_{4_1}|\psi_{4_2})$ take the form:
\begin{equation}
    (\psi_{4_1}|\psi_{4_2})=4\Re \int_{\mathcal{\nu}_0}^{\nu_1}\frac{\Tilde{\psi}_{4_1}(\nu)\Tilde{\psi}^*_{4_2}(\nu)}{S_n(\nu)} d\nu, \label{eq:inner}
\end{equation}
in which the functions $\Tilde{\psi}_4(\nu)$ are the Fourier transforms and $S_n(\nu)$ is the one-sided power spectral density (PSD), that in this context is taken to be flat. The match $\mathcal{F}$ (or faithfulness) is maximized over relative time-shifts and phases between signals.

\begin{figure}[htbp]
\begin{center}
\includegraphics[width=0.45\textwidth]{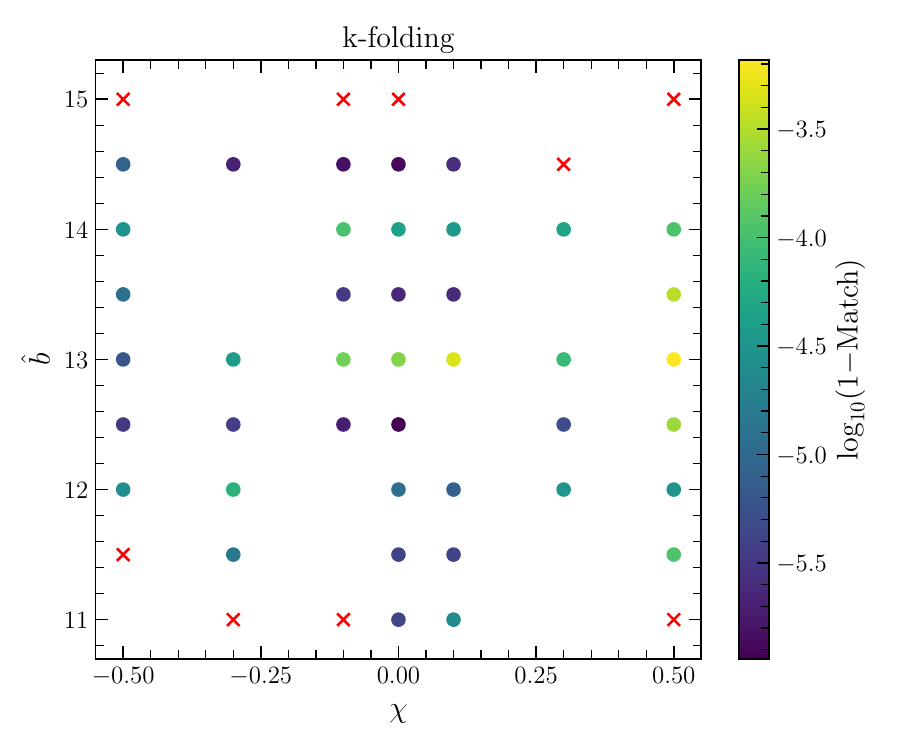}
\caption{Mismatches on the $K$-folding evaluation of the model on the training set (80\% of full dataset), with 7 basis vectors. The red crosses denote waveforms that are outside the convex hull of the fold training set.
\label{fig:kfold_mismatches}}
\end{center}
\end{figure}
\begin{figure}[htbp]
\begin{center}
\includegraphics[width=0.45\textwidth]{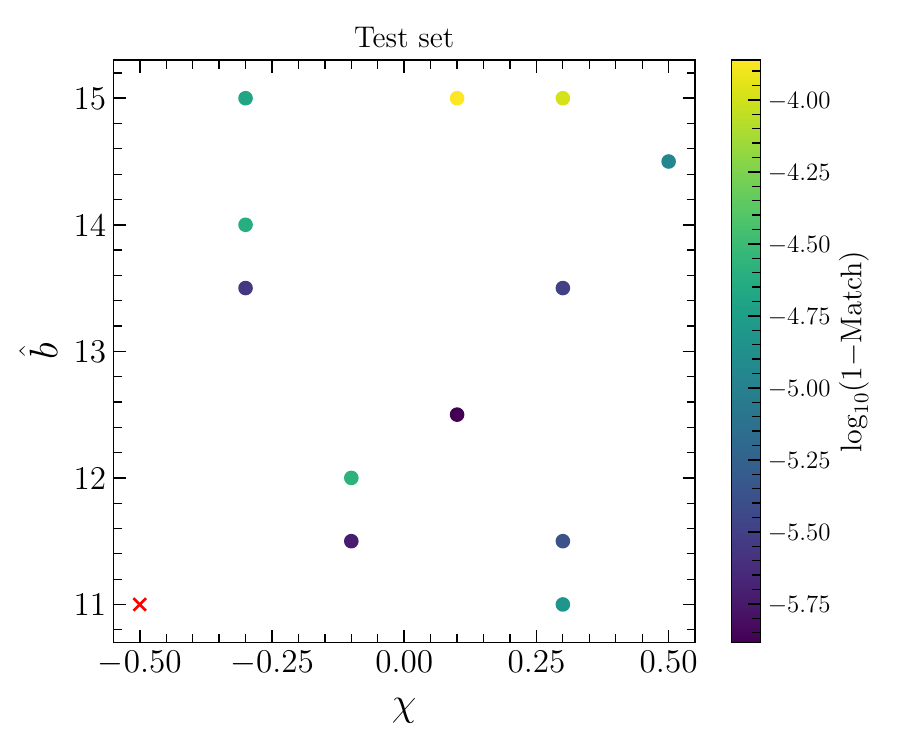}
\caption{Mismatches of the 7-vector surrogate model on the testing set, with 20\% of the full numerical relativity dataset. The red crosses denote waveforms that are outside the convex hull of the model training set.
\label{fig:test_mismatches}}
\end{center}
\end{figure}

As the variability of such waves is quite limited, a basis with three complex vectors seems to be sufficient to describe them. Adding more than three vectors does not significantly improve the accuracy of the model for the current setup. In order to keep a safe margin, we are using a basis of 7 vectors. Fig. \ref{fig:kfold_mismatches} shows the values of the $K$-folding mismatches on the 5 splits of the training set. As expected, the mismatch tends to be larger near the boundaries of the training domain, and also larger for high positive values of the spin parameter $\chi$. On the other hand, in Fig. \ref{fig:test_mismatches} we depict the mismatches of the surrogate (trained on the whole training set) when evaluated on the test set samples.

\section{Injections \label{sec:injections}} 

Given the accuracy of the surrogate model presented in the previous section, we now use it in parameter estimation tasks to understand the measurability of the source parameters . As we will show, the latter can be extremely challenging even if the model is infinitely accurate due to the very subtle imprint of the parameters in the waveform and strong degeneracies present in the parameter space. We inject simulated NR signals corresponding to CHE events into a detector network formed by the LIGO-Handford (H1), LIGO-Livingston (L1) and Virgo (V1) detectors. We do this in zero noise, using the aforementioned power spectral densities of the three detectors at the time of the GW190521 event \cite{GW190521D,GWOSC_GW190521}, converted to their $\psi_4$ versions using the procedure described in \cite{CalderonBustillo:2022cja,CaldernBustillo2023_proca2}. We estimate the parameters of these signals using our surrogate model, within the parameter inference framework \texttt{Parallel Bilby} \cite{Ashton:2018jfp,Smith:2019ucc}. We set uniform priors on the impact parameter $\hat{b}$, the spin and the detector-frame total mass $M$, together with isotropic priors in source orientation and sky-location; and a prior in luminosity distance $d_L$ uniform in co-moving volume. Our injections correspond to non-spinning scattering systems with a detector-frame total mass of $M=60M_\odot$, impact parameters $\hat{b}\in [11,15]$ (CHE regime) separated by steps $\delta 5 = 0.5$. We set the source inclination to $\theta_{\rm JN} = 0.3$ rad and sky-location to $0.8$ rad of right-ascension and $0.3$ rad of declination. This configuration appears to make the signals particularly loud in the L1 detector as compared to the other two. Finally, for all injections, we choose two true distances such that the corresponding optimal SNRs are $\approx$ 15 and 50 in L1, yielding a total of 9 injections for each SNR characterized with a total of 10 parameters.

\subsection{Data pre-processing}

The $\psi_4$ NR data for all our simulations (given in \textit{geometric-solar} units, a.k.a \texttt{Cactus} units: $c=G=M_\odot=1$) is extracted at a radius $R_{\rm ext} = 100M$. The surrogate model is trained with the full set of 63 NR simulations shown in Fig. \ref{fig:param_space} but only taking into account the modes that dominate the outgoing radiation in this kind of system evolution: $(\ell,m)=(2,0)$, $(2,2)$, $(3,2)$ and $(4,4)$. To build the waveforms, each mode is properly time-shifted with respect to $t_{(2,2)}^{\rm peak}$, which is the peak of the most dominant mode, setting it as a reference at $t=0$ \cite{Luna:2024kof}. The other relevant contributions are given by:
\begin{equation}
    \psi_{4,(\ell,-m)}=(-1)^\ell \psi_{4,(\ell,m)}^* \hspace{3mm} (m\neq 0),
\end{equation}
meaning that the complete set of modes conforming the waveform is $\Omega_{(\ell,m)}=\{(2,0),(2,\pm 2),(3,\pm 2),(4,\pm 4)\}$. Hence, the full time-domain NR waveforms injected and the ones recovered after the analysis with the surrogate model take the form:
\begin{equation}
    R_{\rm ext}\psi_4 \simeq \frac{R_{\rm ext}c^4}{GM d_L} \sum_{\Omega_{(\ell,m)}}\psi^{\Omega_{(\ell,m)}}_{4\hspace{4mm}-2} Y_{\Omega_{(\ell,m)}}(\theta_{\rm JN},\phi),\label{eq:phys_wave}
\end{equation}
where $\psi^{\Omega_{(\ell,m)}}_{4} = \psi^{\Omega_{(\ell,m)}}_{4} (\hat{b},\chi)$,  $_{-2}Y_{\Omega_{(\ell,m)}}(\theta_{\rm JN},\phi)$ are the $s=-2$ spin-weighted spherical harmonics, $R_{\rm ext}=100$ is the dimensionless radius of extraction, $d_L$ is the luminosity distance and $M$ is the detector-frame total mass of the system. All the magnitudes are expressed in SI units such that the waves have units of [s$^{-2}$], which is consistent with the fact that $\psi_4=\Ddot{h}$.

\subsection{Data analysis}
The analysis of the waves for the parameter inference has been performed using the Weyl-Scalar $\psi_4$ as described in \cite{CalderonBustillo:2022dph,CalderonBustillo:2022cja}. We attempt to infer a total of 36 parameters, with a particular interest in the impact parameter of the encounter $\hat{b}$ and the individual spins of the BHs $\chi$ (intrinsic parameters). We sample our parameter space with the \texttt{dynesty} \cite{Speagle:2020v2} sampler with the following configuration depending on the SNR:
\begin{table}[h!]
\begin{center}
\caption{Sampler settings as a function of the injected SNR together with its corresponding sampling time.}
\begin{ruledtabular}
%\resizebox{\columnwidth}{!}{%
\begin{tabular}{c c c c c}

Runs     & $\text{SNR}_{\rm Network} $ & $\texttt{n-live}$ & $\texttt{nact}$ & Sampling time     \\ \hline
High SNR & $\approx 54.8$              & $2048$      & $10$             & $\sim 10-15$ days \\
Low SNR  & $\approx 16.7$              & $1024$      & $5$            & $\sim 6-10$ days  \\ 
\end{tabular}%
%}
\label{tab:sampler_settings}
\end{ruledtabular}
\end{center}
\end{table}

We note that the High SNR runs require a more aggressive configuration of the sampler. The strong degeneracies between the parameter space translates to a slow convergence of the sampler with a flat log-likelihood. In particular, we find a strong degeneracy between the intrinsic variables $\hat{b}$ and the source-frame total mass $M^{\rm source}=M/(1+z)$ and the extrinsic variables $d_L$ and $M$. For some of the analysis at High SNR, we get a multi-modal posterior distribution for the total masses $M$ and $M^{\rm source}$ and the impact parameter, precisely due to the poor convergence of the sampler. Proof of that will be shown in the following sections. 

\subsection{Results \label{sec:results}}

\begin{figure*}[!htbp]
\begin{center}
\includegraphics[width=\textwidth]{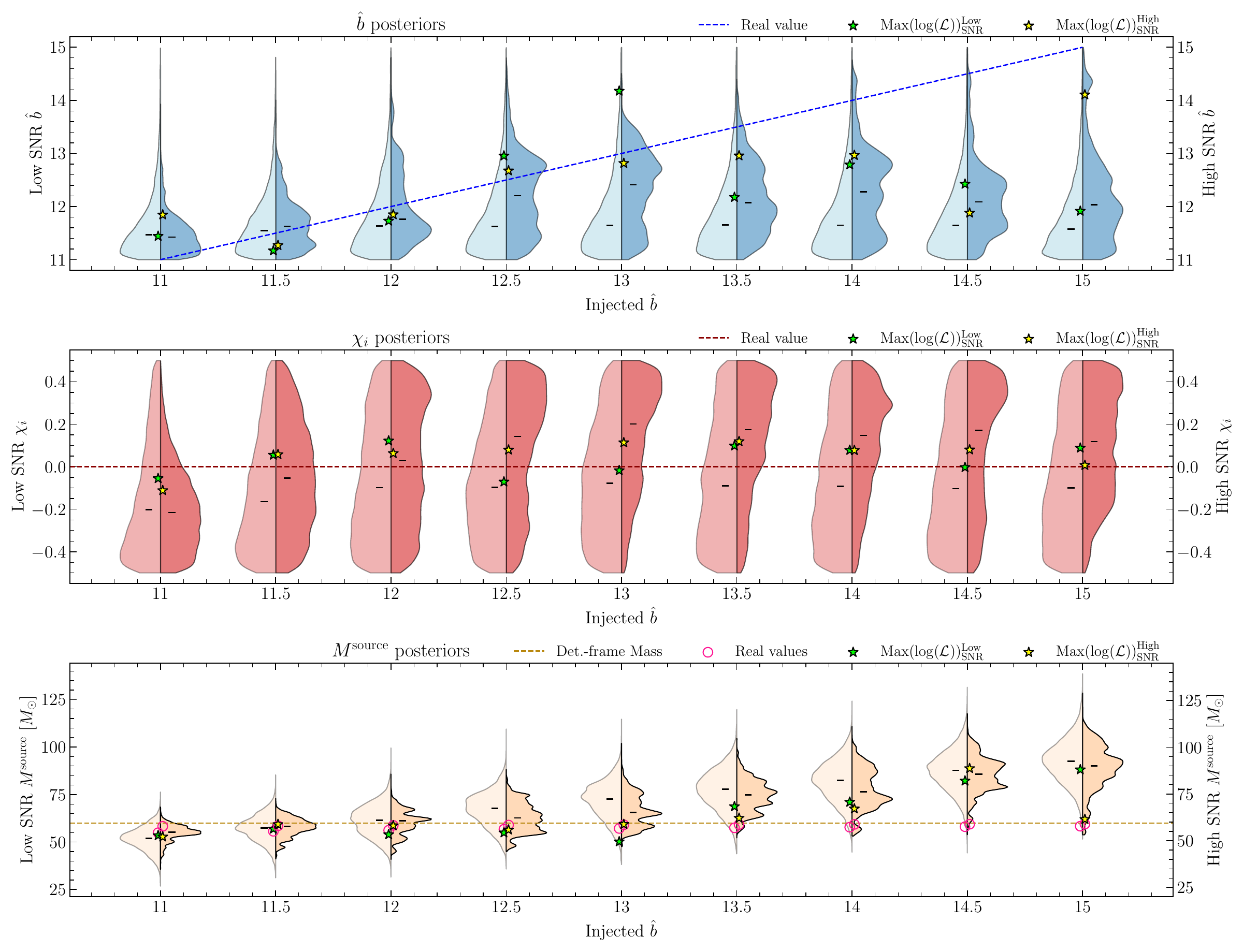}
\caption{Posterior distributions for the impact parameter $\hat{b}$ (top panel), the individual spins $\chi$ (middle panel) and the source-frame total mass $M^{\rm source}$ as a function of the injected $\hat{b}$. On the right part of the \textit{violin} we plot the obtained posterior at High SNR and on the left the posteriors at Low SNR. The black marks correspond to the median of each sample. We also show the values that maximize the log-likelihood in each case as well as the real values (dashed lines for $\hat{b}$ and $\chi$ and circles for $M^{\rm source}$). \label{fig:violin_b_s} }
\end{center}
\end{figure*}

Using the setup described above, we obtain the posterior distributions for both the High and Low SNR cases, for our intrinsic parameters of interest $\hat{b}$, $\chi$ and $M^{\rm source}$. These are shown in Fig. \ref{fig:violin_b_s}. For the case of the impact parameter $\hat{b}$, we always recover a posterior distribution that tends to peak at low values of $\hat{b}$. The reason is that the uniform-comoving volume prior for the luminosity distance strongly favours large source distances and, therefore, intrinsically louder sources \cite{Bustillo2021_Proca1, CaldernBustillo2023_proca2}. This causes the distance to be biased towards larger values than the injected one, at the cost of biasing the intrinsic source parameters to those leading to a larger intrinsic loudness, as it is the case for small $\hat{b}$ values. Similarly, the masses are also biased towards higher values. To compensate for those effects, which are more notorious for Low SNR and as $\hat{b}$ increases (see Appendix \ref{app:posteriors}), the impact parameter is always estimated to be smaller than the injected to match the time-domain amplitude of the signal \eqref{eq:phys_wave}. It was already shown in \cite{Damour:2014afa} that the radiated energy in this kind of events increases significantly as the impact parameter decreases (one order of magnitude difference in our parameter space). 

Regarding the recovery of the spins $\chi$, we obtain posterior distributions that are almost flat for all the runs at both High and Low SNR, meaning that the individual spins are unlikely to be measured. The reason for that is that the spin effects enter at next-to-next to leading order in the Post-Newtonian expansion (1.5 PN) \cite{Morras:2021atg, DeVittori:2014psa, Gopakumar:2011zz, Damour:2004bz, Konigsdorffer:2005sc}. This makes their impact on the waveform being very suppressed by our far-from-merger condition, making their inference being dominated by our Bayesian priors. Nevertheless, the spins play a more important role in other aspects in the physics of our problem, such as the motion of the objects and its corresponding scattering angle (see Appendix \ref{app:spin-effects}). It is worth noticing that the values that maximize the likelihood are always within $\chi \in [-0.15,0.15]$ (reasonably close to the injected value) for both SNRs, unlike for the case of $\hat{b}$, which are spread more randomly. 

\begin{figure}[h!]
\begin{center}
\includegraphics[width=0.45\textwidth]{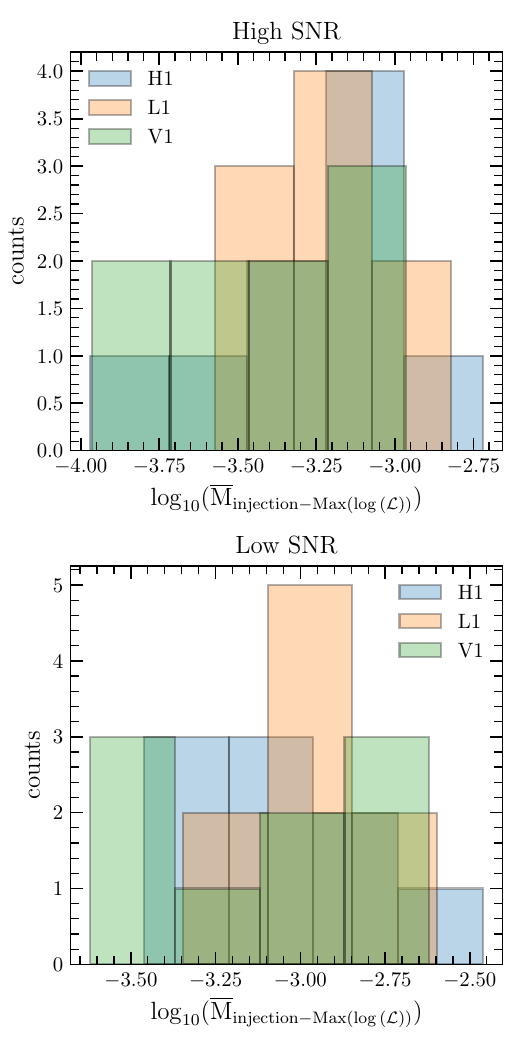}
\caption{Mismatches between the injected waveforms and the ones constructed with the parameters that maximize the log-likelihood at each detector for the High SNR runs (top panel) and the Low SNR runs (bottom panel). \label{fig:mismatches_analysis} }
\end{center}
\end{figure}

Finally, the analysis for the source-frame total mass $M^{\rm source}$ shows the aforementioned poor convergence of these posteriors at High SNR, reflected in their spikiness, due to the strong degeneracy with $\hat{b}$ and $d_L$. Again, the results are better the lower the $\hat{b}$ is. In fact, the posterior distributions for $M^{\rm source}$ are almost identical to the ones for the detector-frame total mass $M$, as the redshifts $z$ considered given the injected distances are small ($\sim 10^{-1}- 10^{-3}$), as shown in the bottom panel from Fig. \ref{fig:violin_b_s} by the dashed line and the circles. Hence, its posteriors do not play a role as important as the ones for $M$ ($\sim 60$), and so $M^{\rm source}\approx M$ (eventhough the redshift and the luminosity distance have a uniform in co-moving volume prior distributions). Similarly to the discussion above, we find that posterior distributions tend to exclude the true injected values for large values of $\hat{b}$. In particular, for the most extreme cases, we see that the true source mass is almost excluded as a consequence of the bias in the luminosity distance we obtain for these systems.

In summary, we find a strong degeneracy between the intrinsic parameters $M^{\rm source}$ and $\hat{b}$ and the extrinsic parameter $d_L$ (and $M$), together with a complete degeneracy in the potential observation of $\chi$. This is something one could already have hinted when analysing the dataset: Fig. \ref{fig:basis_size} shows that our datasets are so simple that can be fairly described with only three vectors in the SVD basis. Hence, in our 10 parameter space there must exist degeneracies. One way to get rid of them might be to increase the SNR to extremely large values, making our analyses way more computationally demanding; or focus on studying other combinations of source parameters, as we will show in the next section. \\

\begin{figure*}[!htbp]
\begin{center}
\includegraphics[width=\textwidth]{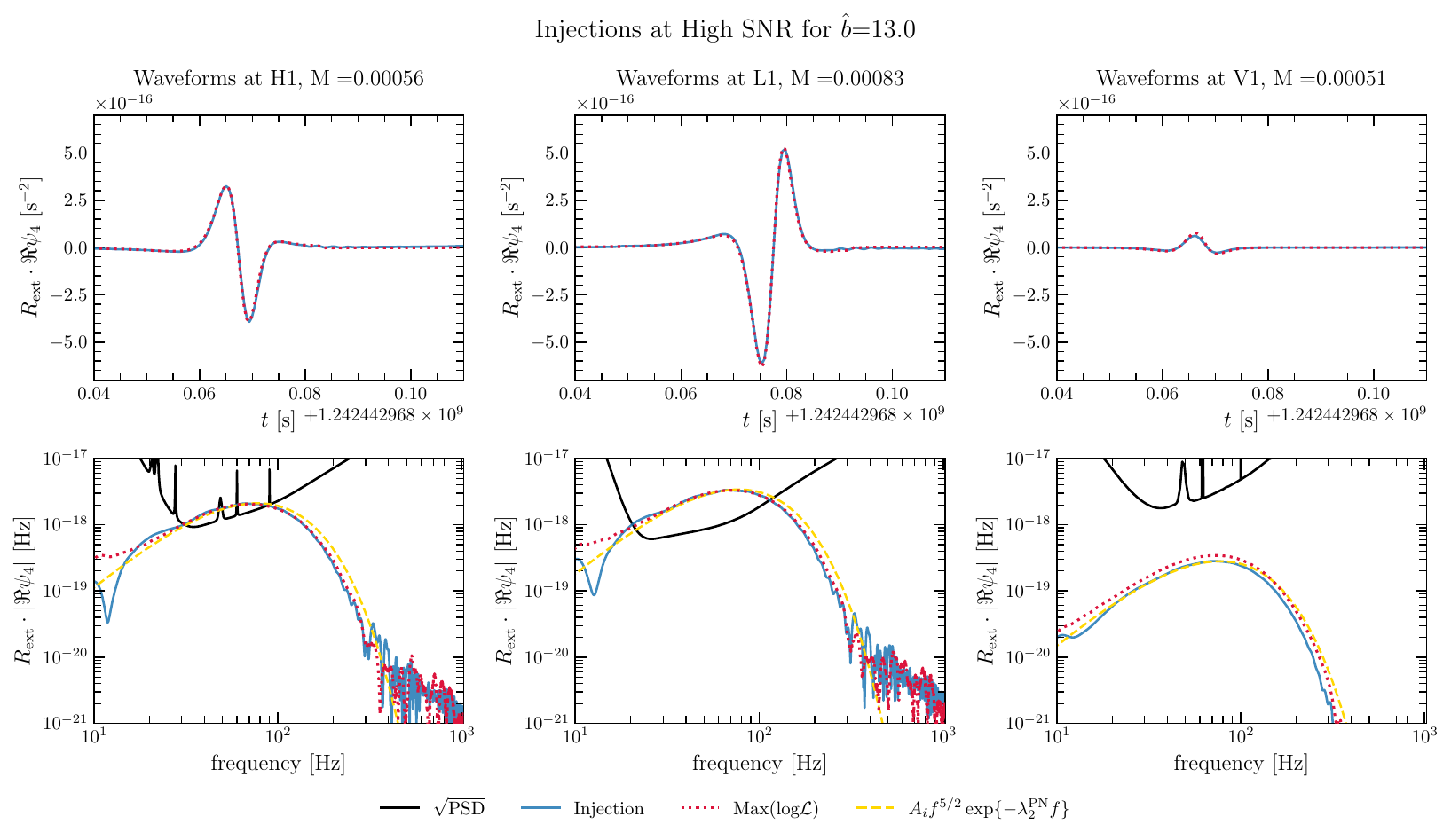}
\caption{Comparison between the injected waves at each detector of the network and the waves constructed with the parameters that maximize the log-likelihood after the analysis for the $\hat{b}=13$ case. We show the real part of the Weyl-Scalar $\psi_4$ both in time domain (top panels) and frequency domain together with the PSD to emphasize the potential observation at each detector (bottom panels), as well as the computed mismatches. For comparison we also show the Newtonian waveform of Eq.~\eqref{eq:h_Newtonian}. The value of $\lambda_2$ is computed using Eq.~\eqref{eq:PN_parameters_surr:l2} to be $\lambda_2 = 0.0323\mathrm{s}$ and the amplitude of the waveform is fitted independently for each interferometer since it will depend on the inclination and sky-location of the event. \label{fig:wf_b13} }
\end{center}
\end{figure*}

Finally, we provide visual proof of the strong challenge that parameter estimation  represents for the type of systems at hand -- in particular measuring $\hat{b}$ and $\chi$ -- comes from the strong degeneracy of the parameter space. 
Fig. \ref{fig:wf_b13} shows the waveforms generated by the parameters that maximise the likelihood, overlaid with those generated with the true injection parameters for the High SNR case with $\hat{b}=13$. The waveforms are indistinguishable by eye both in the time and frequency domains. In addition, to provide a quantitative measurement of such similarity, we show the corresponding mismatches, which are of order $\overline{\rm M} \sim 10^{-4}$ for all detectors. In this same figure (lower panels) we showcase the observability of the injections at each detector by showing the waves in frequency domain together with the corresponding PSDs. %$S_n(\nu)^{1/2}=\sqrt{\rm  PSD}$. 
This signal is sufficiently loud in H1 and L1, but too weak to be measured in V1. In addition, Fig. \ref{fig:mismatches_analysis} shows all the mismatches between the injections and the maximum likelihood recoveries at each detector for all the runs. The mismatches are computed using expressions \eqref{eq:mism} and \eqref{eq:inner} but in this case the weights in the inner product are given by the PSD from each detector in units of $\psi_4$. It is worth noticing how the distribution of the mismatches is broader in H1 and V1 due to the smaller SNR at these detectors, and more peaked in L1, which observes louder signals given our choice of sky-location. 
 
\section{Optimal parameters \label{sec:OP}}

In general, a CHE signal will depend on 17 parameters. These are the two masses $m_1, m_2$, six parameters for the spins $\vec{s}_1, \vec{s}_2$, the eccentricity $e$, impact parameter $b$, inclination $\iota$  and orientation $\Phi$ of the hyperbolic orbit, the right ascension $\alpha$, declination $\delta$, polarization $\psi$, reference time $t_c$ and luminosity distance to the source $d_L$. The surrogate introduced in this paper has significantly less parameters, since it does not depend on the mass ratio $q$, the eccentricity $e$ and the spins $\chi$, which are restricted to be parallel to each other and of equal magnitude.\footnote{While the surrogate does not depend explicitly on the eccentricity, different impact parameters will correspond to different eccentricities since the simulations are done at a constant initial momentum (Eq.~\eqref{eq:P_initial}).} However, this still leads to our CHE model having 10 parameters to describe the morphologically simple signals that can be observed in Fig.~\ref{fig:wf_b13}. Therefore, given the simplicity of the signals, it is natural to expect strong degeneracies between parameters.

Nonetheless, just as in the CBC case, we introduce the chirp mass $\mathcal{M}_c = (m_1 m_2)^{3/5}/(m_1 + m_2)^{1/5}$ as a combination of parameters that can be well measured~\cite{Cutler:1994ys}, as it controls the phase evolution of the signal. Similarly, it is reasonable to expect that in the CHE case there will also be some other combinations of parameters that capture the evolution of prominent features of the signal. Drawing again a parallelism with the CBC case, we will look for these combination of parameters by studying the leading order expressions for the GW emission.

The derivation of the leading order GW emission in the frequency domain can be found in Appendix~\ref{app:NewtonianWaveforms}. We see that it will depend on the parameter $\nu$ defined in Eq.~\eqref{eq:nu_def} as
\begin{align}
    \nu = 2 \pi f \sqrt{\frac{a^3}{G M}} = 6.92 \left(\frac{f}{50 \mathrm{Hz}}\right) \left(\frac{M}{50 M_\odot}\right) \left( \frac{a}{10 R_s}\right)^{3/2} ,
    \label{eq:nu_LIGO}
\end{align}
\noindent where $R_s = 2 G M/c^2$ is the Schwartzshild radius corresponding to the total mass of the system. From Eq.~\eqref{eq:nu_LIGO} we can see that for typical CHEs between two BHs in LIGO, we will have $\nu \gg 1$ and therefore we can use the much simpler approximate expression for the strain of Eq.~\eqref{eq:hphc_Maggiore_f_approx}. We observe that this Newtonian order waveforms are qualitatively consistent with what is shown in Fig.~\ref{fig:wf_b13}, behaving like a power law at low frequency with $|\Tilde{h}| \propto f^{1/2} \rightarrow |\Tilde{\psi}_4| \propto f^{5/2}$ and exponentially decaying at high frequency. Such a simple waveform, with characteristic strain
\begin{equation}
    \Tilde{h}_c(f) = \lambda_1 \sqrt{f} \exp\{-\lambda_2 f\} \, ,
    \label{eq:h_Newtonian}
\end{equation}
\noindent will have two measurable intrinsic parameters, the overall amplitude $\lambda_1$ and the argument of the exponential $\lambda_2$ which will control the peak frequency of the strain, located at $f_\mathrm{peak} = 1/(2 \lambda_2)$. From Eq.~\eqref{eq:hphc_Maggiore_f_approx} we observe that these two parameters will be 
\begin{subequations}
\label{eq:PN_parameters}
\begin{align}
\lambda_1 & =  \frac{8 \pi \eta}{c^4 e^2 d_L} \left[(e^2 - 1) G M a\right]^{5/4} \, , \label{eq:PN_parameters:l1} \\
\lambda_2 & = \left(\sqrt{e^2-1} - \arctan{\sqrt{e^2 - 1}}\right) 2 \pi \sqrt{\frac{a^3}{G M}} \, , \label{eq:PN_parameters:l2}
\end{align}
\end{subequations}
\noindent where $\eta = m_1 m_2/(m_1 + m_2)^2$ is the symmetric mass ratio and we have used that $\sec^{-1}(e) = \arctan\sqrt{e^2 - 1}$. As mentioned previously, the surrogate introduced in this paper will not depend on all the parameters introduced here, since the simulations are done at constant mass ratio $(m_1 = m_2 = M/2)$ and initial momentum $P = 0.11456439 M c$. Therefore, we want to take into account these restrictions on the parameters of Eq.~\eqref{eq:PN_parameters} and write them in terms of the variables we use in parameter estimation, i.e. the dimensionless impact parameter $\hat{b}$ at the initial time, the total mass $M$ and the luminosity distance $d_L$. 

To do this, we note that the impact parameter of the NR simulation $\hat{b}$ will not be the same as the Newtonian impact parameter $\hat{b}_\infty$, defined when the two black holes are at infinite separation. However, the two of them will be related by the conservation of the dimensionless angular momentum

\begin{equation}
    \hat{L} = \frac{c}{G M} L = \hat{P} \hat{b} = \hat{P}_\infty \hat{b}_\infty \, ,
    \label{eq:b_of_binf}
\end{equation}

\noindent where we have defined the dimensioless momenta as 

\begin{equation}
\hat{P} \equiv \frac{P}{\mu c} \, .
\label{eq:hatP_def}
\end{equation}

In our simulations $\hat{P}= 0.45825756$ and $\hat{P}_\infty$ can be computed using that

\begin{equation}
    \hat{P}_\infty = \sqrt{\frac{2(E_\mathrm{ADM} - M c^2)}{\mu c^2}} = 0.4252 \, ,
    \label{eq:P_infty}
\end{equation}

\noindent where we have used that for our simulations $E_\mathrm{ADM} = 1.0226 M c^2$. Using then the expression of the Newtonian energy of Eq.~\eqref{eq:Hiperb_params_LO:a}, given by

\begin{align}
    E =\frac{1}{2} c^2 \hat{P}_\infty^2 = \frac{G M}{2 a} \, ,
    \label{eq:E_Newtonian}
\end{align}

\noindent we can solve for the  semi-major axis of the system $a$ as

\begin{equation}
    a  = \frac{G M}{c^2 \hat{P}^2} = 2.77 R_s \left( \frac{0.4252}{\hat{P}_\infty} \right)^2 \, , \label{eq:a_Newtonian}
\end{equation}

\noindent and using that for a hyperbolic orbit the impact parameter is given by 

\begin{equation}
b = a \sqrt{e^2 - 1} \longrightarrow \hat{b}_\infty = \frac{c^2 b}{G M} = \frac{c^2 a}{G M} \sqrt{e^2 - 1} \, ,
\label{eq:b_Newtonian}
\end{equation}

\noindent we can use Eq.~\eqref{eq:a_Newtonian} to find the Newtonian eccentricity $e$ of the system to be

\begin{align}
\sqrt{e^2 -1} = \hat{P}_\infty^2 \hat{b}_\infty = \hat{P}_\infty \hat{P} \hat{b}  = 2.14 \left( \frac{\hat{P}_\infty \hat{P}}{0.1949} \right) \left(\frac{\hat{b}}{11} \right) \, . \label{eq:e_Newtonian}
\end{align}

In Eq.~\eqref{eq:e_Newtonian} we observe that the Newtonian eccentricities explored by the surrogate are very large, being between $e \in (2.37, 3.09)$. Using Eqs.~(\ref{eq:a_Newtonian},\ref{eq:e_Newtonian}), we can simplify Eq.~\eqref{eq:PN_parameters} in terms of the surrogate parameters as
\begin{subequations}
\label{eq:PN_parameters_surr}
\begin{align}
\lambda_1 & = 8 \pi \eta \frac{c}{d_L} \frac{\left(\hat{P} \hat{b} G M/c^3 \right)^{5/2}}{1 + (\hat{P}_\infty \hat{P} \hat{b})^2} \, , \label{eq:PN_parameters_surr:l1} \\
\lambda_2 & = 2 \pi \frac{G M }{c^3 \hat{P}_\infty^3}\left(\hat{P}_\infty \hat{P} \hat{b} - \arctan\{\hat{P}_\infty \hat{P} \hat{b}\} \right) \, . \label{eq:PN_parameters_surr:l2}
\end{align}
\end{subequations}

The leading order (Newtonian) GW emission of Eq.~\eqref{eq:h_Newtonian} computed using the value of $\lambda_2$ of Eq.~\eqref{eq:PN_parameters_surr:l2} is shown in Fig.~\ref{fig:wf_b13}. We observe a good agreement with the NR-based surrogate model, further strengthening the case that this parametrization is a good basic description of the signal.

\subsection{Results: degeneracy study}

\begin{figure*}[!ht]
\begin{center}
\includegraphics[width=\textwidth]{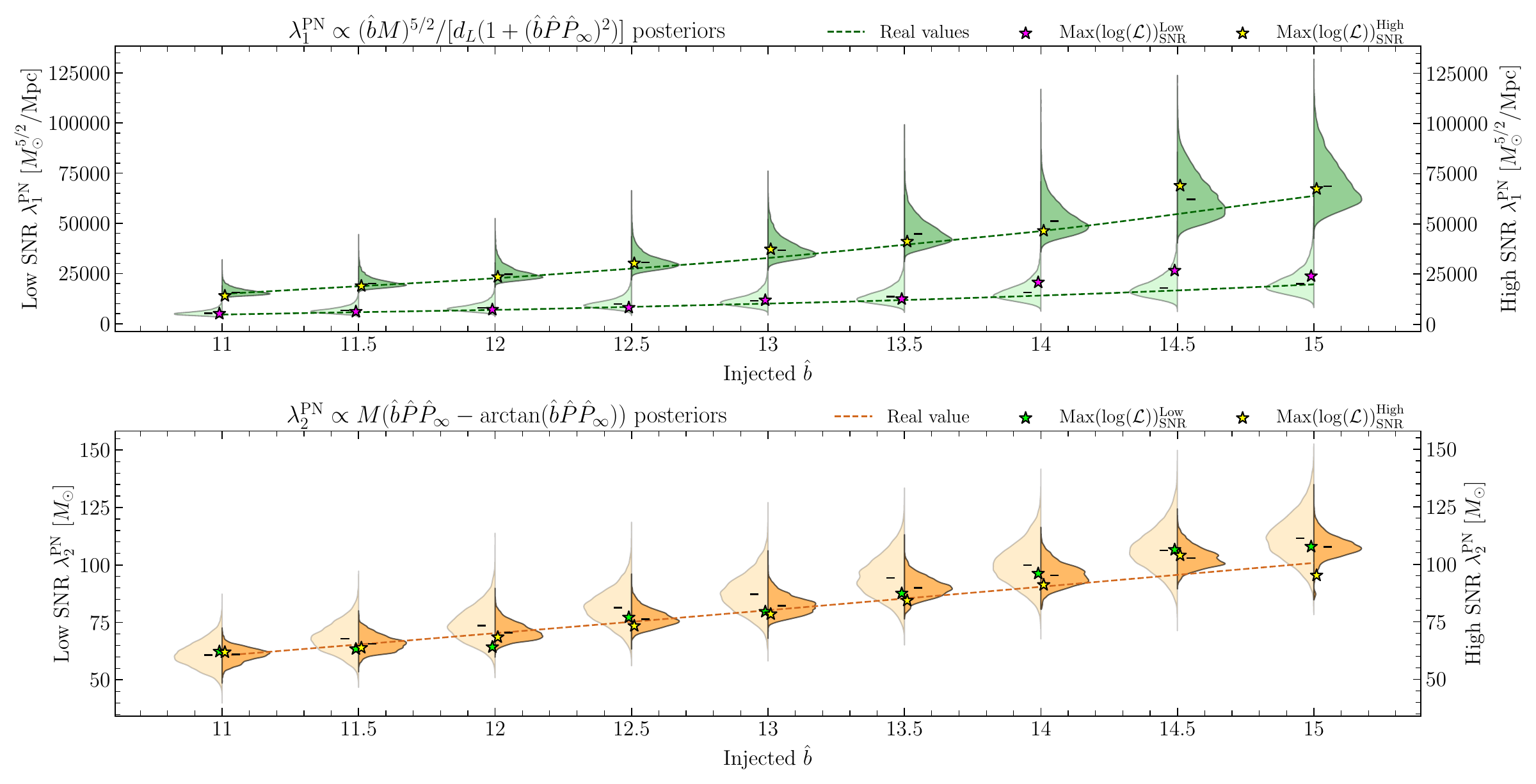}
\caption{Posterior distributions for the parameters $\lambda_1$ (top panel) and $\lambda_2$ (bottom panel) (\ref{eq:PN_parameters_surr:l1},\ref{eq:PN_parameters_surr:l2}) that describe the Newtonian waveform in frequency domain \eqref{eq:h_Newtonian} as a function of the injected $\hat{b}$. On the right part of the \textit{violin} we plot the obtained posterior at High SNR and on the left the posteriors at Low SNR. The black marks correspond to the median of each sample. We also show the values that maximize the log-likelihood in each case as well as the real values (dashed lines). \label{fig:violin_pn} }
\end{center}
\end{figure*}
\begin{figure*}[!htbp]
\begin{center}
\includegraphics[width=\textwidth]{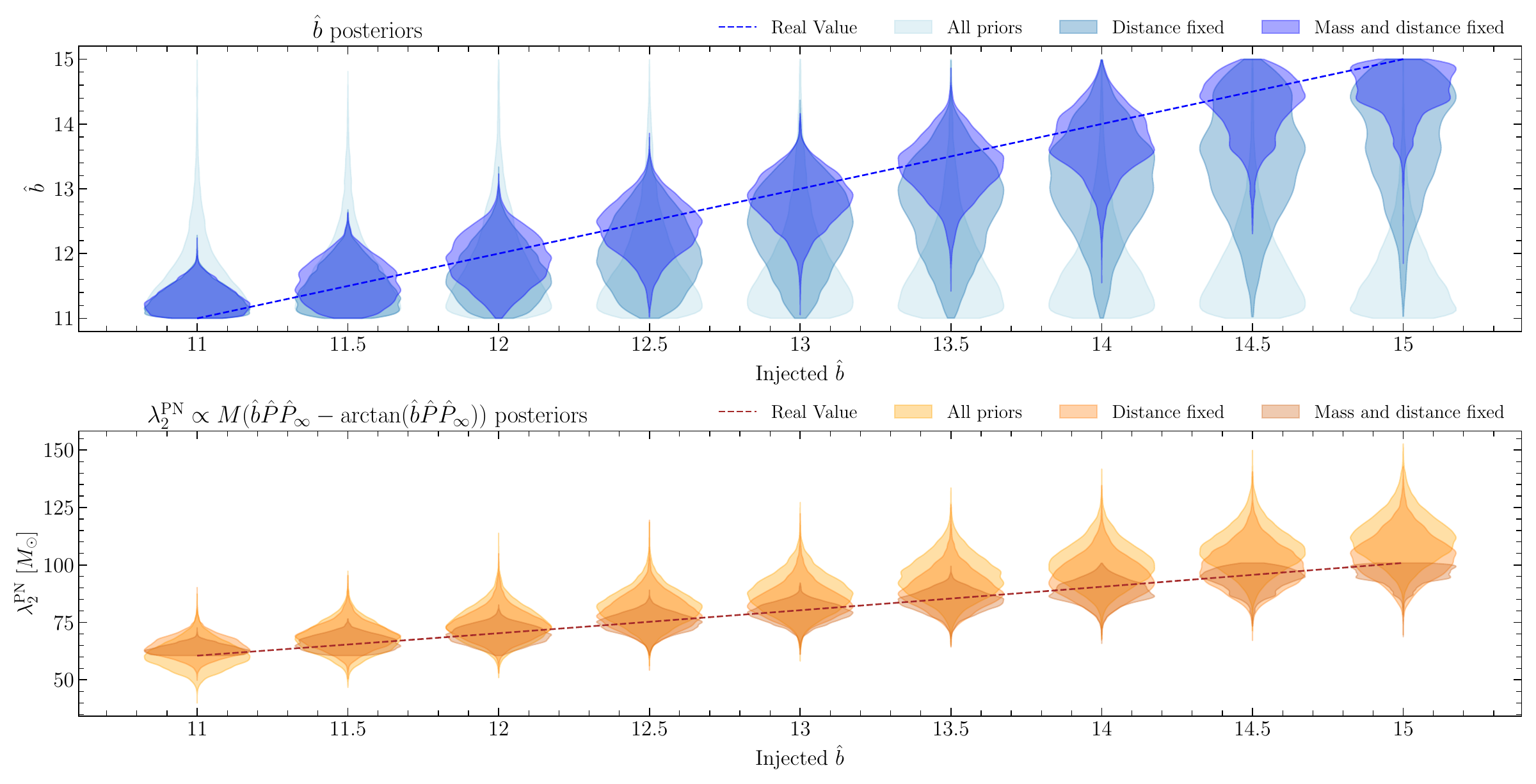}
\caption{Posterior distributions for the impact parameter $\hat{b}$ (top panel) and the PN parameter $\lambda_2$ \eqref{eq:PN_parameters_surr:l2} (bottom panel) as a function of the injected $\hat{b}$ at Low SNR. We show the comparison between the posteriors obtained when considering all prior distributions in the analysis with the cases in which relevant priors are fixed to the injected value, together with the real values (dashed lines). \label{fig:violin_b_deg} }
\end{center}
\end{figure*}

With these new \textit{optimal parameter choices} at hand \eqref{eq:PN_parameters_surr:l1}-\eqref{eq:PN_parameters_surr:l2}, we now study their measurability. %we can study its potential observation by combining the posterior distributions obtained in Sec. \ref{sec:injections}. 
The corresponding posterior distributions are shown in Fig. \ref{fig:violin_pn} as a function of the injected impact parameter $\hat{b}$. It is straightforward to see how the distributions for both $\lambda_1$ and $\lambda_2$ are more Gaussian, or at least display a distinguishable peak, also at High SNR. This fact allows us to confirm that the multi-modalities and poor convergence of the sampler shown in Fig. \ref{fig:violin_b_s} are caused by the physics of the sources we study. On top of that, we can affirm that the parameter related to the amplitude of the Newtonian waveform $\lambda_1$ is accurately inferred, as it depends on the variables that are more strongly degenerated $\hat{b}$, $M$ and $d_L$. On the other hand, $\lambda_2$ is recovered better for low $\hat{b}$ and High SNR, as it lacks dependence on $d_L$, thus causing a higher deviation for larger $\hat{b}$, which is precisely where the effect of the degeneracies is more notorious. Nevertheless, we can see how, in general for both parameters, the waveforms that maximize the likelihood are consistently close to the injected values represented by the dashed lines, which are constructed by cubic interpolation of the injected values as a function of $\hat{b}$, as they clearly appear to be smooth functions in this parameter space.  

The top panel has its posteriors centered at different values due to their dependence on the true $d_L$ of the injection, chosen such that the SNR is maintained constant for different $\hat{b}$. It is worth noticing how the corresponding distributions have a larger variance for High SNR, as a consequence of $\sim d_L^{-1}$ dependence of the parameter (the posteriors at High SNR are always less dispersed than for Low SNR). This effect generates the convenient paradox of recovering more precisely the injected values the lower the SNR is. The posteriors in the bottom panel behave in the opposite way, constraining better $\lambda_2$ the higher the SNR is. Some of the posteriors for High SNR case still present some irregularities due to the poor convergence given the strong degeneracies, as shown in Fig. \ref{fig:violin_b_s}, specially for the detector-frame total mass and the impact parameter. \\

These results illustrate the fact that one might be able to measure some general features of the gravitational waves produced by these type of events with the drawback of ignoring part of the relevant intrinsic information that characterizes them, even with the current ground-based detectors, given the promising quality of the posterior distribution at Low SNR shown in Fig. \ref{fig:violin_pn}. For the sake of verifying this last statement, we show in Fig. \ref{fig:violin_b_deg} the breaking of the degeneracy at Low SNR when one provides the sampler with the real information about some of the degenerated parameters by removing the prior distributions of such parameters. We visualize how we are able to recover satisfactorily the impact parameter $\hat{b}$ (top panel) and $\lambda_2$ (bottom panel) when the total mass and luminosity distance priors are removed and instead we pass to the sampler the injected values. We can also qualitatively observe how by just fixing the distance, the posterior distributions are substantially improved, emphasizing a stronger degeneracy between $d_L$ and $\hat{b}$ than between $M$ and $\hat{b}$, explained by the dominance of the uniform in co-moving volume prior distribution for the luminosity distance.   

\section{Conclusions}

    We have performed 63 NR simulations of spinning and non-spinning CHEs at a fixed energy $E_{\rm ADM}\approx1.023M$, covering the range of impact parameters $\hat{b}\in[11,15]$ and individual spins $\chi\in[-0.5,0.5]$ such that all the configurations result in an open orbit. We use the $\psi_4$ extracted from these simulations to train a surrogate model for CHEs presented in Sec. \ref{sec:surr}, 
    including the leading mode $(\ell,m)=(2,\pm 2)$ mode and the sub-leading $(2,0)$, $(3,\pm 2)$ and $(4,\pm  4)$  modes. 
    In the range of parameters covered by our study, we observe that only three (complex) vectors in the SVD basis suffice to describe
    the waveforms in the surrogate model, showing the relative simplicity of this data. 
    We test the performance of the model by computing the mismatches between the NR waveforms following a K-fold cross validation procedure, obtaining values for the unfaithfulness below $10^{-3}$, 
    see Fig. \ref{fig:kfold_mismatches} and Fig. \ref{fig:test_mismatches}. 
   %
   %The singular value decomposition employed to build the surrogate model reveals the low complexity of the CHEs's waveforms in our parameter space, which they appear to be able to be described with only three vectors of the SVD basis. %This fact already hints that the parameter inference task might be challenging.
    
    %After validating the performance of the surrogate model, 
    %One of the central motivations of this work has been to carry out  using the employ it to carry out parameter inference tasks. 

    Following model validation, we carry out parameter inference studies on numerically simulated signals using our surrogate model.
    We overcome the lack of a robust method to obtain the strain from the $\psi_4$ data for non-circular binaries by using the $\psi_4$-based framework described in \cite{CalderonBustillo:2022dph,CalderonBustillo:2022cja}.
    %together with surrogate model trained with $\psi_4$ data. 
    %
    The analysis of the waveforms injected to the LVK network that we present in Sec. \ref{sec:injections} shows that even with the high accuracy of our surrogate model and SNRs as large as $\sim 50$, there are strong degeneracies among the parameters that define the waveforms. 
    This can be concluded noting that the posterior distributions of intrinsic parameters (e.g impact parameter, individual spins and source-frame total mass) 
    %that contain the relevant physical information about such events 
    %\sout{are off with respect to the injected data, specially as the injected impact parameter increases; the weaker the gravitational field is, the lower impact the intrinsic parameters have on the waveform and the larger is the dominance of the distance priors.}
    are not only broad but also centered around values which do not coincide with the injected ones. This effect becomes stronger with larger impact parameters due to the increasing weakness of the gravitational field. %\jc{I think that the explanation of how one concludes the existence of degeneracies is not really needed in the conclusions}
    \footnote{This is very similar to why one can constrain much better spin angles if measured close to merger.}
    In particular, the strong bias induced by the dominance of the prior in co-moving volume for the luminosity distance, which favours more distant sources (as we showed in Appendix \ref{app:posteriors}), causes the exclusion of the true injected values for the impact parameter and the source mass for weak sources, biasing these towards values characteristic of louder sources.
    Moreover, we obtain  posterior distributions for the spins which are almost flat, showing that measuring such parameter will be extremely challenging %difficulties of measuring the  injected value.
    \footnote{This might be expected given the sub-leading effect of spin in a PN expansion, see e.g. \cite{Morras:2021atg, DeVittori:2014psa, Gopakumar:2011zz, Damour:2004bz, Konigsdorffer:2005sc}. Nevertheless, we show in Appendix \ref{app:spin-effects} that spin has a noticeable impact on the scattering angles. }
    %which means that it's going to be very unlikely to measure it. 
    %
    Despite the challenge that measuring the aforementioned parameter represents, in Fig. \ref{fig:wf_b13} we note that the waveforms which maximize the likelihood match very well the NR injections. Once again, this is a hallmark of the degeneracies of the waveforms in parameter space. 
    %within our parameter space and the high accuracy of the surrogate.
    Finally, we stress that physical signals similar to the ones considered here lie within the sensible frequency band of the current ground-base detectors, so that its observation is currently technologically feasible, but unfortunately also very astrophysically unlikely \cite{Bini:2023gaj}. 
    %by comparing the frequency domain representation of the by showing the frequency domain representation of the waves, we compare them with the PSDs at each detector, showing that they lie within the sensible frequency band of the current ground-base detector. 
    
    %This last statement is reinforced by the study developed 
    
    Given this observational feasibility, in Sec. \ref{sec:OP} we study combinations of parameters in which these degeneracies could be ameliorated, similarly to the chirp mass for compact binary coalescences.
    To do so, we derive the Newtonian waveform in frequency domain for hyperbolic orbits \eqref{eq:h_Newtonian}, which reveals two possible observables: the overall amplitude and the argument of the exponential damping factor which controls the peak frequency, given by Eqs.(\ref{eq:PN_parameters_surr:l1},\ref{eq:PN_parameters_surr:l2}), both combinations of the more strongly degenerated parameters $\hat{b}$, $M$ and $d_L$. 
    Notably, we find that the posteriors for these parameters are more sharply defined and are more compatible with the true injected values (specially $\lambda_1$, as it depends on the aforementioned three most degenerated parameters), see Fig. \ref{fig:violin_pn}.
    We also verify that if one removes the priors of some of these degenerated parameters in the injection, the posteriors of the remaining ones are consistent with the injected true values, see Fig. \ref{fig:violin_b_deg}.

    Summing up, we conclude that although the individual values of the 
    intrinsic parameters could be difficult to measure even with high SNR events, certain combinations of them can be properly inferred.
    We leave for future work the assessment of how informative these parameters are to constrain astrophysical and cosmological predictions. 
    
    %Furthermore, the question now would be if such combinations make sense astrophysically and cosmologically speaking.
    %In conclusion, it is going to be extremely tough to accurately identify the intrinsic parameters that define these systems, probably even with the third generation detectors and extremely loud signals. 
    %
    %This work suggests that only certain combinations of parameters are expected to be properly inferred, with the drawback of ignoring the intrinsic physical information that define these systems. 
    
    A natural continuation of this work is to expand the coverage of parameter space of the surrogate model considering varying mass ratios and initial energies. Moreover, it would be interesting to explore the transition to merging binaries in the stronger field regime.
    
    %improve the surrogate model by not only increasing the current range of the parameters considered (in order to study the CHE-to-merger transition) but also the parameter space itself, training it to cover other intrinsic features of these systems such as the mass ratio and the initial energy. 

    %[Finally we encourage the NR, GR Approximants, Waveform and Parameter Inference communities to delve deeper in this topic, by performing more NR simulations and improving the approximants that describe CHEs, as some other groups are currently doing \cite{Rettegno:2023ghr, Albanesi:2024xus}. ] \TA{I'm not a fan of these kind of sentences but up to you!}

\section{Acknowledgements}

JF acknowledges support by the EU Horizon under ERC Consolidator Grant, no. InspiReM- 101043372. 
The work of TA is supported in part by the Ministry of Science and Innovation (EUR2020-112157, PID2021-125485NB-C22, CEX2019-000918-M funded by MCIN/AEI/10.13039/501100011033), and by AGAUR (SGR-2021-01069).
%
% Raimon Postdoc Grants:
RL acknowledges financial support provided by APOSTD 2022 post-doctoral grant CIAPOS/2021/150, funded by Generalitat Valenciana / Fons Social Europeu ``L'FSE inverteix en el teu futur''. RL also acknowledges financial support provided by the Individual CEEC program 2023.06381.CEECIND of 2023, funded by the Portuguese Foundation for Science and Technology (Funda\c{c}\~{a}o para a Ci\^{e}ncia e a Tecnologia).
%
% Raimon Spaninsh Projects:
This work is supported by the Spanish Agencia Estatal de Investigaci\'on (grant PID2021-125485NB-C21) funded by MCIN/AEI/10.13039/501100011033 and ERDF ``A way of making Europe'', and the Generalitat Valenciana (grant CIPROM/2022/49).
%
% Raimon Portuguese Projects:
This work is supported by the Center for Research and Development in Mathematics and Applications (CIDMA) through the Portuguese Foundation for Science and Technology (FCT - Funda\c{c}\~{a}o para a Ci\^{e}ncia e a Tecnologia) through projects: 
UIDB/04106/2020 (DOI identifier \url{https://doi.org/10.54499/UIDB/04106/2020}); 
UIDP/04106/2020 (DOI identifier \url{https://doi.org/10.54499/UIDP/04106/2020}); 
PTDC/FIS-AST/3041/2020 (DOI identifier \url{http://doi.org/10.54499/PTDC/FIS-AST/3041/2020}); and 
2022.04560.PTDC (DOI identifier \url{https://doi.org/10.54499/2022.04560.PTDC}).
JCB is partially funded by a fellowship from ``la Caixa'' Foundation (ID100010474) and from the European Union's Horizon2020 research and innovation programme under the Marie Skodowska-Curie grant agreement No 847648. The fellowship code is LCF/BQ/PI20/11760016. JCB is also supported by the research grant PID2020-118635GB-I00 and the a Ram\'{o}n y Cajal Fellowship RYC2022-036203-I from the Spain-Ministerio de Ciencia e Innovaci\'{o}n. 
RL and JCB acknowledge further support by the European Horizon Europe staff exchange (SE) programme HORIZON-MSCA2021-SE-01 Grant No. NewFunFiCO-101086251. 
GM, SJ and JGB acknowledge support from the Spanish Research Project PID2021-123012NB-C43 [MICINN-FEDER], and the Centro de Excelencia Severo Ochoa Program CEX2020-001007-S at IFT. SJ is supported by the FPI grant PRE2019-088741 funded by MCIN/AEI/10.13039/501100011033. GM acknowledges support from the Ministerio de Universidades through Grant No. FPU20/02857.
The EinsteinToolkit simulations were performed at MareNostrum4 and MareNostrum5 at the Barcelona Supercomputing Center (Grants No. AECT-2022-2-0006 and No. FI-2024-2-0012) and at the Lluis Vives cluster for scientific calculations at University of Valencia\footnote{\url{https://www.uv.es/uvweb/servei-informatica/ca/serveis/}}.
We acknowledge computational resources provided by the CIT cluster of the LIGO Laboratory and supported by National Science Foundation Grants PHY-0757058 and PHY0823459; and the support of the NSF CIT cluster for the provision of computational resources for our parameter inference runs. This material is based upon work supported by NSF’s LIGO Laboratory which is a major facility fully funded by the National Science Foundation.

%%%%%%%%%%%%%%%%%%%%%%%%%%%%%%%%%%%%%%%%%%%%%%%%%%%
\appendix 

\section{Details on consistency checks: scattering angles and convergence test}
\label{app:consistency}

\begin{figure}[!htbp]
\begin{center}
\includegraphics[width=0.45\textwidth]{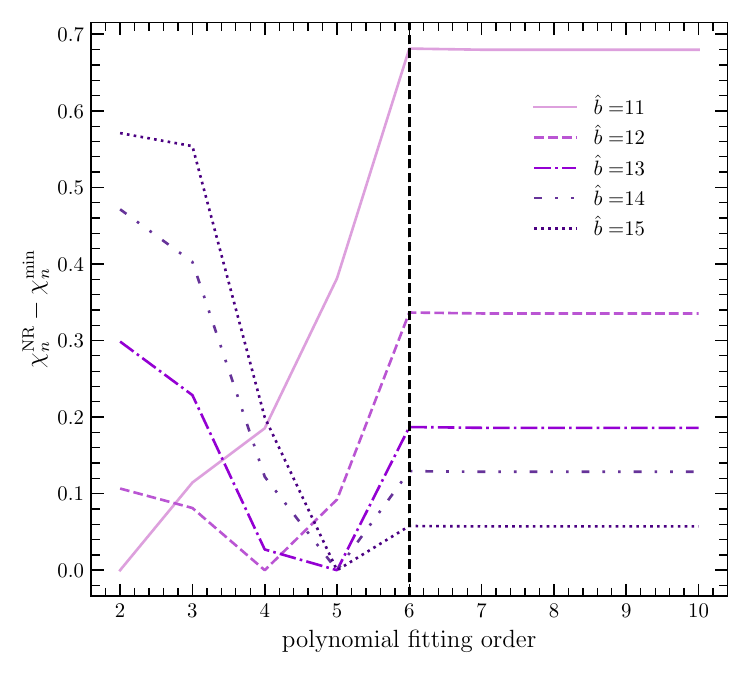}
\caption{\label{fig:scat_fits} We represent the differences $\chi^{\rm NR}_{n}-\chi^{\rm min}_{n}$ to show the fact that a polynomial of order $n=6$ is enough to provide a consistent result for the scattering angle.}
\end{center}
\end{figure}

We compare our results for the gauge invariant scattering angle $\chi^{\rm NR}$ of the spin-less configurations with the computations published in \cite{Damour:2014afa}. We follow their approach to compute them by extrapolating the tracks of the BHs. This is not the most orthodox method because the positions of the objects are not gauge invariant, although it has been shown in previous research that the gauge choosing effects on the tracking should be small, based on the agreement found between the analytical and numerical approaches \cite{Damour:2014afa,Hopper:2022rwo,Rettegno:2023ghr,Damour:2022ybd,Albanesi:2024xus}. This discussion suggests that one should be conservative when it comes to assign an error to the final results for the scattering angles. We proceed as detailed in \cite{Albanesi:2024xus,Damour:2014afa}, computing the scattering angles by extrapolating the tracks of the BHs using $1/r$ polynomials of different orders starting from $n=2$: we make a change to the relative motion between the objects in the polar coordinates $(r,\varphi)$, allowing us to extrapolate the incoming and outgoing angles to infinity, leading to:
\begin{equation}
    \chi^{\rm NR} = \varphi_{\rm out}^{\infty} - \varphi_{\rm in}^\infty -\pi \label{eq:chi_NR}
\end{equation}
The polynomial fits are performed with a least-squares method that uses a singular-value decomposition (SVD). It is found that a polynomial of order $n=6$ is enough to give a consistent result with higher order fits using the SVD detailed in \cite{Albanesi:2024xus}, as we show in Fig. \ref{fig:scat_fits}.

Hence, we can assign this value as the result. The error is conservatively estimated by the difference between the maximum and minimum angles computed using the different order polynomials extrapolations $\Delta\chi^{\rm NR}_{\rm fit}=\chi^{\rm max}_{n}-\chi^{\rm min}_{n}$. With this approach, we find a remarkable agreement with the computations made in \cite{Damour:2014afa}, coinciding within less than a $0.7\%$. We represent the data in TABLE \ref{tab:chi_comparison}.  \\

\begin{figure}[!htbp]
\begin{center}
\includegraphics[width=0.45\textwidth]{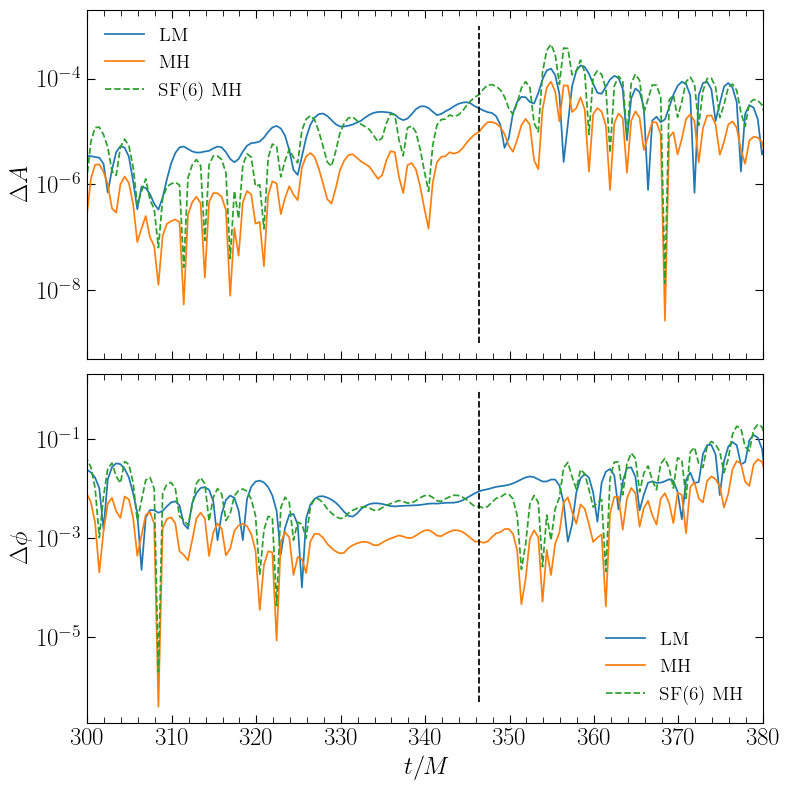}
\caption{\label{fig:convb11} Convergence test for the $\hat{b} = 11$ configuration. We show the amplitude (top) and phase (bottom) differences for consecutive resolutions. The dashed vertical line marks merger time. The dashed green lines correspond to the Medium-High lines scaled by the scaling factor (SF) for order 6.}
\end{center}
\end{figure}

\begin{table}[!h]
\caption{\label{tab:chi_comparison} Our results for the gauge-invariant scattering angles of the spinless configurations, and those presented in \cite{Damour:2014afa}, in degrees.}
\begin{center}
\begin{ruledtabular}
\begin{tabular}{c c c} 
$\hat{b}$ & $\chi^{\rm NR}$ & $\chi^{\rm NR}_{{\rm Damour \, et \, al}}$ \\
\hline
\hline
11.0 & $152.6\pm 0.5$ & $152.0(1.3)$  \\
12.0 & $120.9\pm 0.6$ & $120.7(1.5)$ \\
13.0 & $101.9\pm 0.8$ & $101.6(1.7)$ \\
14.0 & $88.8\pm 1.1$ & $88.3(1.8)$ \\
15.0 & $78.9 \pm 1.3$ & $78.4(1.8)$ 
\end{tabular}
\end{ruledtabular}
\end{center}
\end{table}
We have performed the simulation with $\hat{b} = 11$ at various resolutions, varying the grid spacing at the coarsest level, $dx = 8, 6, 4 M$. We show the amplitude and phase differences for consecutive amplitudes as a function of time in Fig \ref{fig:convb11}. 

We have checked for convergence of various orders as follows. For a general order of convergence $r$, we define the scaling factor
\begin{equation}
    SF(r) = \frac{dx_L^r - dx_M^r}{dx_H^r - dx_M^r},
\end{equation}
\noindent where $dx_{R_i}$ is the coarsest grid separation of resolution $R_i =$ Low, Medium, High.
The approximate correspondence of $(A_M - A_H) SF(r) \approx (A_L - A_M)$ and similarly for the phases indicates good convergence of this order. 
Our results are consistent with convergence of order 6, which is relatively high although smaller than the order of spatial differentiation in our runs which is 8. 

\section{Impact parameter and SNR effect on the posterior distributions \label{app:posteriors}}

We can qualitatively describe the deviation of the posterior distributions with respect to the injected values by computing the relative offset of the samples means: 
\begin{equation}
    \text{Offset}(\mathcal{O}) = \frac{\overline{\mathcal{O}}}{\mathcal{O}_{\rm inj}}-1 \label{eq:offset}
\end{equation}
\begin{figure}[h!]
%\begin{center}
\includegraphics[width=0.45\textwidth]{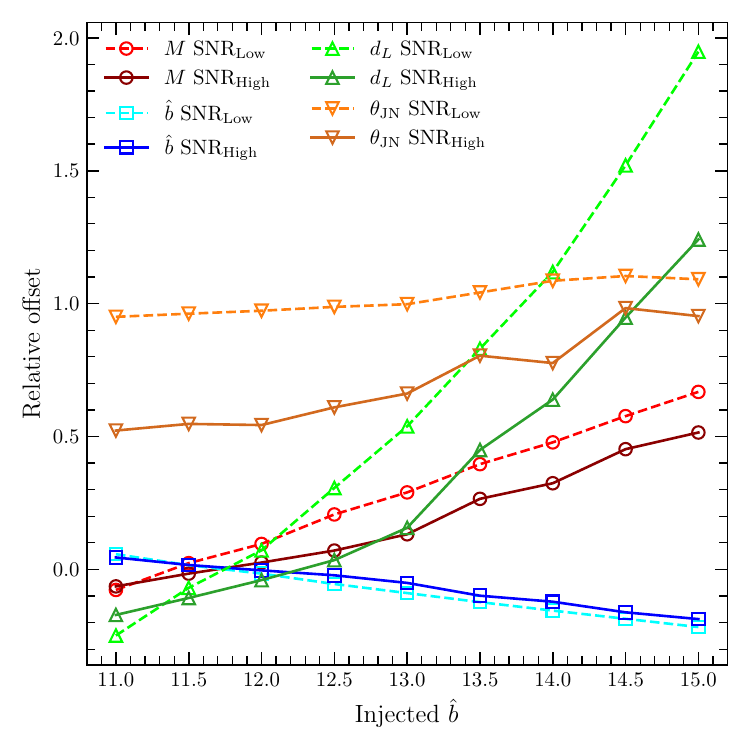}
\caption{\label{fig:offsets} Relative offsets of the means with respect to the injected values of the total mass, impact parameter, luminosity distance and inclination for both High and Low SNR.}
%\end{center}
\end{figure}

This is represented in Fig. \ref{fig:offsets} for the detector-frame mass $M$, impact parameter $\hat{b}$, the luminosity distance $d_L$ and the inclination $\theta_{\rm JN}$ at Low and High SNR. In Fig. \ref{fig:offsets} we exemplify the dominance of the prior in co-moving volume for the luminosity distance as $\hat{b}$ increases, together with the preference to assign larger masses. We clearly see how these effects are more notorious the lower the SNR is. We also show the case for the inclination $\theta_{\rm JN}$, an intrinsic parameter which is inferred almost equally for all the range in $\hat{b}$, but is always overestimated. Larger inclinations mean lower intrinsic emitted amplitude, which indicates another reason to consistently underestimate $\hat{b}$ to compensate for that. 

\section{Spin effects \label{app:spin-effects}}

\begin{figure}[h!]
%\begin{center}
\includegraphics[width=0.45\textwidth]{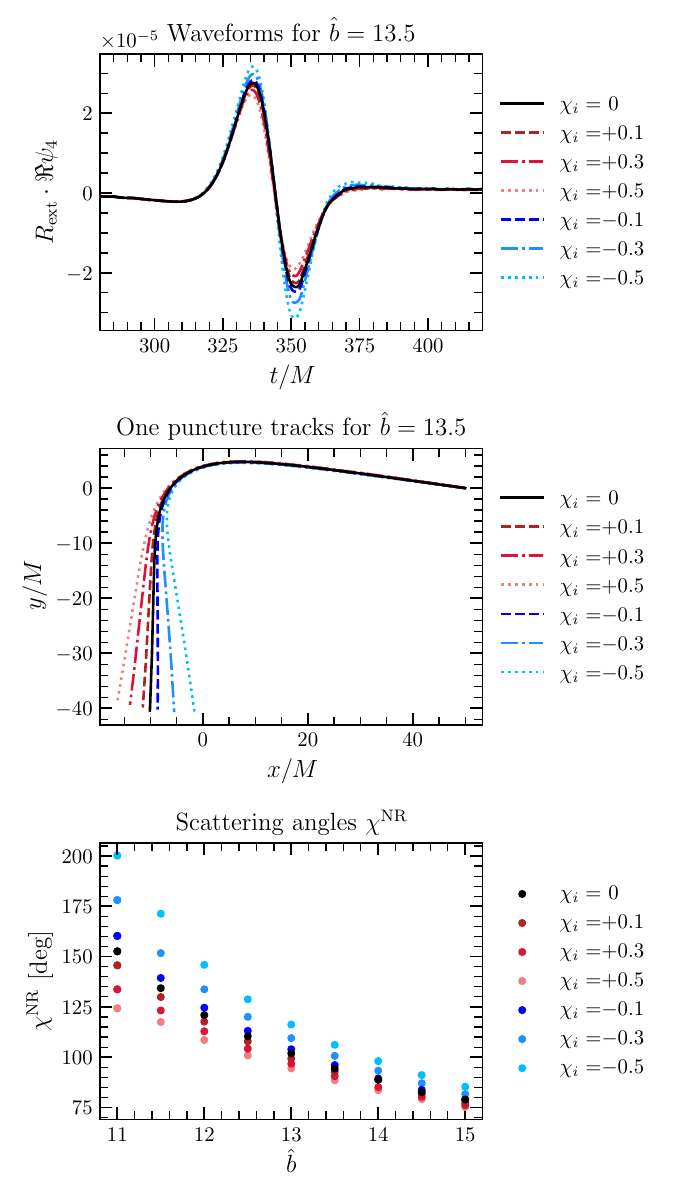}
\caption{\label{fig:spin-effects} We show the impact of the individual spins $\chi_i$ on the waveform (top panel), on the trajectory of one of the BHs (middle panel) and on the gauge invariant scattering angle $\chi^{\rm NR}$ (bottom panel). The error bars for the scattering angles are small and thus are not shown.}
%\end{center}
\end{figure}

One of the main goals of this research has been to attempt to infer the intrinsic parameters that define a CHE event. We showed in Fig. \ref{fig:violin_b_s} that this might not be an easy task due to the strong degeneracy within the parameter space, even for signals as loud as $\text{SNR}\sim 50$. In particular, the obtained posterior distributions for the spins of the BHs are practically flat, meaning that their inference is dominated by the Bayesian priors. We have observed though, that the spins have a relevant impact on the physics of this kind of event. The PN effects of the spin have been studied exhaustively in the literature \cite{Morras:2021atg, DeVittori:2014psa, Gopakumar:2011zz, Damour:2004bz, Konigsdorffer:2005sc}. Although they enter at 1.5 PN order for both the motion and the gravitational radiation, the NR data unveils the fact that the impact for the former is more notorious than for the later, specially as $\hat{b}$ increases. We show this comparison in Fig. \ref{fig:spin-effects}, where we plot the raw $\psi_4$ NR data (top panel), the track of one of the BHs (middle panel) and the scattering angles (bottom panel) for our 63 simulations in the $\chi\in[-0.5,0.5]$ and $\hat{b}\in[11,15]$ regime. We can appreciate how for the example case of $\hat{b}=13.5$, the spin effects are qualitatively more notorious in the motion rather than in the waveform. Consequently, the gauge invariant scattering angles computed as described in the Appendix \ref{app:consistency} seem to be considerably more dependent on the spins. Nevertheless, these results suggest that there will still exist a degeneracy between the spins, the impact parameter and the scattering angles if one improves the surrogate model presented in this paper by training it with the scattering angles data or if in the future one is able to compute them from the waves with a robust method.

\section{Newtonian Waveforms}
\label{app:NewtonianWaveforms}

In this appendix we will derive the leading order (Newtonian) gravitational wave polarizations emitted by a CHE. In Newtonian mechanics, the Kepler problem for an unbound system has the well known solution
\begin{subequations}
\label{eq:Hiperb_LO}
\begin{align}
t & = \nu_0 (e \sinh{\xi} - \xi) \, , \label{eq:Hiperb_LO:t}\\
r & = a(e \cosh{\xi} - 1) \, , \label{eq:Hiperb_LO:r} \\
\Phi & = 2 \arctan\left({\sqrt{\frac{e+1}{e-1}}}\tanh{\frac{\xi}{2}}\right) \, , \label{eq:Hiperb_LO:Phi}
\end{align}
\end{subequations}
\noindent which are hyperbolic orbits with eccentricity $e$, semimajor axis $a$ and timescale $\nu_0$. These properties of the orbit can be written in terms of the conserved quantities (i.e. energy $E$ and angular momentum $L$) as
\begin{subequations}
\label{eq:Hiperb_params_LO}
\begin{align}
 a & = \frac{G M}{2 E} \, , \label{eq:Hiperb_params_LO:a}\\
 e & = \sqrt{1 + \frac{2 E L^2}{(G M)^2}} \, , \label{eq:Hiperb_params_LO:e} \\
 \nu_0 & = \frac{G M}{(2 E)^{3/2}} = \sqrt{\frac{a^3}{G M}} \, , \label{eq:Hiperb_params_LO:nu_0}
\end{align}
\end{subequations}
\noindent where $M = m_1 + m_2$ is the total mass of the binary. Since the angular momentum is conserved, the orbit can be assumed to happen in a plane. If we choose $\vec{L} = L \hat{z}$, it will be contained in the $x-y$ plane and using Eq.~\eqref{eq:Hiperb_LO} we have
\begin{subequations}
\label{eq:cartesian_xi}
\begin{align}
 x(\xi) & = r \cos{\Phi} = a (e - \cosh{\xi})  \, , \label{eq:cartesian_xi:x}\\
 y(\xi) & = r \sin{\Phi} =  a \sqrt{e^2 - 1} \sinh{\xi} \, , \label{eq:cartesian_xi:y}\\
 z(\xi) & = 0 \, , \label{eq:cartesian_xi:z}
\end{align}
\end{subequations}
\noindent where we have chosen the $x$-axis to be along the periastron line. To compute the gravitational waves (GWs) we will use Eq.~(3.72) from Ref.~\cite{Maggiore:2007ulw}, taking into account that the orbit takes place in the $x-y$ plane:
\begin{subequations}
\label{eq:hphc_Maggiore}
\begin{align}
 h_+ = \frac{G}{c^4 d_L} \Big[& \ddot{M}_{11} \left( \cos^2 \phi - \sin^2 \phi \cos^2 \theta \right) \nonumber \\
+ & \ddot{M}_{22} \left( \sin^2 \phi - \cos^2 \phi \cos^2 \theta \right) \nonumber \\
- & \ddot{M}_{12} \sin{2 \phi} \left(1 + \cos^2 \theta \right) \Big] \, , \label{eq:hphc_Maggiore:hp}\\
 h_\times = \frac{G}{c^4 d_L} \Big[& (\ddot{M}_{11} - \ddot{M}_{22}) \sin{2 \phi}\cos\theta \nonumber \\
 + & 2\ddot{M}_{12} \cos{2 \phi} \cos\theta\Big] \, , \label{eq:hphc_Maggiore:hc}
\end{align}
\end{subequations}
\noindent where $d_L$ is the luminosity distance between the CHE and the detector, $(\phi, \theta)$ are the polar angles of the direction of emission of the FW with respect to the $(x,y,z)$ frame in which the problem was solved, and $M_{ij}$ is defined as: 
\begin{equation}
    M_{i j} = \int \d^3 \vec{x} x_i x_j T^{00}(t_r, \vec{x}) \, .
    \label{eq:M_ij_def}
\end{equation}
Since we are considering two point masses in the Newtonian aproximation, we can take $T^{00}(t_r, \vec{x}) = \mu \delta(\vec{x} - \vec{r}(t_r))$, and thus:
\begin{equation}
    M_{i j} = \mu r_i r_j \, .
    \label{eq:M_ij_r_i_r_j}
\end{equation}
Using the orbit in Eq.~\eqref{eq:cartesian_xi}, and ignoring constant terms that do not contribute to the GW emission, we obtain:
\begin{widetext}
\begin{equation}
    M_{i j} = \frac{1}{2} \mu a^2
    \begin{pmatrix}
        \cosh{2 \xi} - 4 e \cosh{\xi} & \sqrt{e^2 -1} \left( 2 e \sinh{\xi} - \sinh{2 \xi} \right) & 0\\
         \sqrt{e^2 -1} \left( 2 e \sinh{\xi} - \sinh{2 \xi} \right) & (e^2 -1) \cosh{2 \xi} & 0 \\
        0 & 0 & 0
    \end{pmatrix}
    \, .
    \label{eq:M_ij_xi_LO}
\end{equation}
\end{widetext}
We could compute the derivatives with respect to time of $M_{i j}$ by using Eqs.~\eqref{eq:Hiperb_LO} together with the chain rule. It will be more interesting however to compute the waveforms directly in the frequency domain. To do this we will want to compute the Fourier transform of Eq.~\eqref{eq:hphc_Maggiore}, i.e.
\begin{equation}
    \widehat{h}_a(f) = \int_{-\infty}^{\infty} \d t \, h_a(t) \rme^{\rmi 2 \pi f t} = \int_{-\infty}^{\infty} \d t \, h_a(t) \rme^{\rmi \omega t} \, .
    \label{eq:hphc_Fourier}
\end{equation}
Because of this, we will have to compute the Fourier transform of $\ddot{M}_{i j}$:
\begin{equation}
    \widehat{\ddot{M}}_{i j} = - \omega^2 \widehat{M}_{i j} \, .
    \label{eq:M_ij_Fourier}
\end{equation}
To compute the Fourier transform of $M_{i j}$, given in Eq.~\eqref{eq:M_ij_xi_LO}, we will use the following expressions of Ref.~\cite{Garcia-Bellido:2017knh}:
\begin{subequations}
\label{eq:Fourier_hiperb}
\begin{align}
 \widehat{\sinh{\xi}} & = - \frac{\pi}{\omega}\left( \frac{1}{e} H^{(1)}_{\rmi \nu}(\rmi \nu e) \right)  \, , \label{eq:Fourier_hiperb:shx}\\
 \widehat{\cosh{\xi}} & = - \frac{\pi}{\omega}\left( H^{(1)\prime}_{\rmi \nu} (\rmi \nu e) \right) \, , \label{eq:Fourier_hiperb:chx}\\
 \widehat{\sinh{2\xi}} & = - \frac{\pi}{ \omega}\left( \left(\frac{4}{e^2} - 2 \right) H^{(1)}_{\rmi \nu}(\rmi \nu e) + \frac{4 \rmi }{\nu e} H^{(1)\prime}_{\rmi \nu} (\rmi \nu e) \right)  \, , \label{eq:Fourier_hiperb:sh2x}\\
 \widehat{\cosh{2\xi}} & = - \frac{\pi}{ \omega}\left( \frac{4}{e} H^{(1)\prime}_{\rmi \nu}(\rmi \nu e) + \frac{4 \rmi}{\nu e^2} H^{(1)}_{\rmi \nu} (\rmi \nu e) \right)  \, , \label{eq:Fourier_hiperb:ch2x}
\end{align}
\end{subequations}
\noindent where $H^{(1)}_{\rmi \nu}(\rmi \nu e)$ is the Hankel function of the first kind~\cite{10.5555/1098650} and $\nu$ is a normalized frequency given by:
\begin{equation}
    \nu = \nu_0 \omega = 2 \pi \nu_0 f \, .
    \label{eq:nu_def}
\end{equation}
Using these results, the Fourier transform of $\ddot{M}_{i j}$ is given by:
\begin{equation}
    \widehat{\ddot{M}}_{i j} = 2 \pi \mu \sqrt{G M a} \, \widehat{m}_{i j}
    \label{eq:M_ij_of_m_ij}
\end{equation}
\noindent where
\begin{subequations}
\label{eq:m_ij_Fourier_Hankel}
\begin{align}
 \widehat{m}_{11} = &  \frac{\rmi }{e^2} H^{(1)}_{\rmi \nu}(\rmi \nu e) - \nu \left( e - \frac{1}{e} \right) H^{(1)\prime}_{\rmi \nu}(\rmi \nu e) \, , \label{eq:m_ij_Fourier_Hankel:11} \\
 \widehat{m}_{22} = &  \frac{\rmi }{e^2} H^{(1)}_{\rmi \nu}(\rmi \nu e) + \frac{\nu}{e} H^{(1)\prime}_{\rmi \nu}(\rmi \nu e) \, , \label{eq:m_ij_Fourier_Hankel:22} \\
 \widehat{m}_{12} = &  \nu \left( 1 - \frac{1}{e^2} \right) H^{(1)}_{\rmi \nu}(\rmi \nu e) - \frac{\rmi }{e} H^{(1)\prime}_{\rmi \nu}(\rmi \nu e) \, . \label{eq:m_ij_Fourier_Hankel:12}
\end{align}
\end{subequations}
And the gravitational waves are then given by:
\begin{subequations}
\label{eq:hphc_Maggiore_f}
\begin{align}
 \Tilde{h}_+(f)  = 2 \pi \frac{G \mu \sqrt{G M a}}{c^4 d_L} \Big[ & \widehat{m}_{11} \left( \cos^2 \phi - \sin^2 \phi \cos^2 \theta \right) \nonumber \\
+ & \widehat{m}_{22} \left( \sin^2 \phi - \cos^2 \phi \cos^2 \theta \right) \nonumber \\ 
- & \widehat{m}_{12} \sin{2 \phi} \left(1 + \cos^2 \theta \right) \big] \, , \label{eq:hphc_Maggiore_f:hp}\\
 \Tilde{h}_\times(f) = 2 \pi \frac{G \mu \sqrt{G M a}}{c^4 d_L} \big[ &  (\widehat{m}_{11} - \widehat{m}_{22}) \sin{2 \phi}\cos\theta \nonumber \\
 + & 2\widehat{m}_{12} \cos{2 \phi} \cos\theta\big] \, . \label{eq:hphc_Maggiore_f:hc}
\end{align}
\end{subequations}
In Ref.~\cite{Garcia-Bellido:2017knh}, we can also find the large frequency approximation of the Hankel functions appearing in Eq.~\eqref{eq:m_ij_Fourier_Hankel}, i.e.
\begin{subequations}
\label{eq:HankelApprox}
\begin{align}
& H_{\rmi \nu}^{(1)}(\rmi \nu e) \! = \! -\rmi  \sqrt{\frac{2}{\pi \nu \sqrt{e ^2-1}}} \rme^{\nu \left(\sec ^{-1}(e )-\sqrt{e ^2-1}\right)}\! \left[ \! 1 \! + \! \mathcal{O} \! \! \left( \! \frac{1}{\nu} \! \right) \!\right] \! , \\
& H_{\rmi \nu}^{(1)}{}'(\rmi \nu e) = \sqrt{\frac{2 \sqrt{e^2-1}}{\pi \nu e^2}} \rme^{\nu \left(\sec ^{-1}(e )-\sqrt{e ^2-1}\right)} \! \left[ \! 1 \! + \! \mathcal{O} \! \! \left( \! \frac{1}{\nu} \! \right) \!\right] \! . 
\end{align}
\end{subequations}
Substituting these approximations in the expressions for the GW polarizations of Eq.~\eqref{eq:hphc_Maggiore_f} and keeping leading order terms in $\nu$, we obtain
\begin{subequations}
\label{eq:hphc_Maggiore_f_approx}
\begin{align}
 \Tilde{h}_+(f)  & = - \frac{1 + \cos^2\theta}{2} \rme^{-2 \rmi  \phi} \, \Tilde{h}_c (f)  \, , \label{eq:hphc_Maggiore_f_approx:hp}\\
 \Tilde{h}_\times(f) & = - \rmi  \cos{\theta} \rme^{-2 \rmi  \phi} \, \Tilde{h}_c (f) \, , \label{eq:hphc_Maggiore_f_approx:hx} \\
 \Tilde{h}_c (f) & = 8 \pi  \frac{G \mu \sqrt{\sqrt{G M \ell^5} f}}{c^4 e^2 d_L}  \rme^{\nu \left(\sec^{-1}(e )-\sqrt{e^2-1}\right)}, \label{eq:hphc_Maggiore_f_approx:hc}
\end{align}
\end{subequations}
\noindent where $\ell$ is the semilatus rectum, defined as
\begin{equation}
    \ell = (e^2 - 1) a.
    \label{eq:semilatus_rectum}
\end{equation}

\bibliography{refs.bib, references.bib}
\end{document}